\documentclass[10pt,conference]{IEEEtran}

\usepackage{color}

\ifx\pdfoutput\undefined
\usepackage{graphicx}
\graphicspath{{eps_figs/}}
\else
\usepackage[pdftex]{graphicx}
\graphicspath{{pdf_figs/}}
\fi

\usepackage{latexsym,amssymb,amsmath,ifthen,afterpage,cite}

\newcommand{\Fig}[1]{Fig.~\ref{#1}}

\newcommand{\eqdef}{\stackrel{\scriptscriptstyle\bigtriangleup}{=} }

\newcommand{\R}{\mathbb{R}}

\newcommand{\B}{{\mathcal{B}}}

\newcommand{\calX}{\mathcal{X}}

\newcommand{\calU}{\mathcal{U}}

\newcommand{\X}{\bf{X}}
\newcommand{\x}{{\bf x}}

\newcommand{\E}{\operatorname{E}}
\newcommand{\V}{\operatorname{Var}}

\newcounter{examplecntr}
{\begin{trivlist}\small\item[]\refstepcounter{examplecntr}%
 {\bfseries Example~\theexamplecntr%
  \ifthenelse{\equal{#1}{}}{}{ (#1)}.
}}%
{\end{trivlist}}

\newcounter{theoremcntr}
{\begin{trivlist}\item[]\refstepcounter{theoremcntr}%
{\bfseries Theorem~\thetheoremcntr%
  \ifthenelse{\equal{#1}{}}{}{ (#1)}.
}}%
{\hfill$\Box$\end{trivlist}}

\newcommand{\pos}[2]{\makebox(0,0)[#1]{#2}}

\let\oldbibliography\thebibliography
\renewcommand{\thebibliography}[1]{%
  \oldbibliography{#1}%
  \setlength{\itemsep}{2.2pt}%
}

\begin{document}
\DeclareGraphicsExtensions{.pdf}

\title{An Importance Sampling Scheme on Dual Factor Graphs.~I.~Models in a Strong External Field
} 

\author{Mehdi Molkaraie \\
\tt{mehdi.molkaraie@alumni.ethz.ch}
}

\maketitle 
 
\begin{abstract} 
We propose an importance sampling scheme to estimate the 
partition function of the two-dimensional 
ferromagnetic Ising model and the two-dimensional 
ferromagnetic $q$-state Potts model, both in 
the presence of an external magnetic field. 
The proposed scheme operates 
in the dual Forney factor graph and is capable of efficiently  
computing an estimate of the partition function under a wide range of
model parameters. In particular, we consider models that are 
in a strong external magnetic field.
\end{abstract} 
 
\section{Introduction} 
 
In~\cite{MoLo:ISIT2013}, the authors showed that for 
two-dimensional (2D) Ising
models, at low temperature  
Monte Carlo methods mix
much faster on the dual Forney factor graph than on the original factor graph.
Monte Carlo methods on the dual factor graph were also
proposed in~\cite{MoLo:ISIT2013}
to estimate of the partition function of 2D Ising models (with 
constant or with spatially varying couplings) in 
the absence of an external magnetic field.

In the absence of an 
external field, the exact value of the partition function of 
2D Ising models 
with constant coupling was first calculated by 
Onsager~\cite{Onsager:44},~\cite[Chapter 7]{Baxter07}.
However, the 2D Ising model in an arbitrary non-zero external 
field and the three-dimensional (3D) Ising model have remained 
unsolved~\cite{Welsh:90,Cipra:00}. 

In general, quantities of interest in statistical physics, e.g., the partition 
function and the mean magnetization of 2D models,
can be estimated 
using Markov chain Monte 
Carlo methods~\cite{HH:64,Neal:proinf1993r,BH:10,LoMo:IT2013}.
At low temperatures, however, 
Monte Carlo methods usually suffer from critical slowing down. 
It is well known that at a certain 
critical temperature, the 2D 
ferromagnetic
Ising model undergoes a phase transition;
below this temperature, variables (spins) have long-range dependencies and
Monte Carlo methods (based on single spin-flips) do 
not mix rapidly~\cite{BH:10}. 

We propose an importance sampling algorithm~\cite{HH:64,Neal:proinf1993r}
which can be used to compute
the partition function of models
with pairwise interactions. In this paper, we are mainly concerned with computing the 
partition function
of finite-size 2D ferromagnetic Ising models
and $q$-state Potts models~\cite{Potts:52}, 
when the models 
are under the influence of an external field.
In our numerical experiments,
we will also consider 3D 
ferromagnetic Ising models.
The importance sampling 
scheme operates on the dual of the Forney factor graphs representing the models. 
Our numerical results show that the scheme performs
well in a wide range of model parameters.

It must be emphasized that, unlike well-known 
algorithms, e.g., Gibbs sampling~\cite{GG:srgd1984} and 
the Swendsen-Wang
algorithm~\cite{SW:87}, the proposed scheme does not suggest 
a method to draw samples according to the Boltzmann distribution on factor graphs,
as sampling is done in the dual domain. 

The rest of the paper is organized as follows. In Section~\ref{sec:Ising},
we review the Ising model
and graphical model representations in terms of 
Forney factor graphs. 
Dual Forney factor graphs and 
the normal factor graph duality theorem
are discussed in Section~\ref{sec:NFGD}.
The importance sampling algorithm on the
dual Forney factor graph is described in Section~\ref{sec:IS}. 
In Section~\ref{sec:Potts},
we briefly discuss generalizations to the $q$-state Potts model.
Numerical experiments are reported in Section~\ref{sec:Num}.

\section{The Ising Model in an External Magnetic Field}
\label{sec:Ising}

\begin{figure}[t]
\setlength{\unitlength}{0.89mm}
\centering
\begin{picture}(81,66)(0,0)
\small
\put(0,60){\framebox(4,4){$=$}}
 \put(4,60){\line(4,-3){4}}
 \put(8,54){\framebox(3,3){}}
\put(4,62){\line(1,0){8}}        \put(8,63){\pos{bc}{$X_1$}}
\put(12,60){\framebox(4,4){}}
\put(16,62){\line(1,0){8}}
\put(24,60){\framebox(4,4){$=$}}
 \put(28,60){\line(4,-3){4}}
 \put(32,54){\framebox(3,3){}}
\put(28,62){\line(1,0){8}}       \put(32,63){\pos{bc}{$X_2$}}
\put(36,60){\framebox(4,4){}}
\put(40,62){\line(1,0){8}}
\put(48,60){\framebox(4,4){$=$}}
 \put(52,60){\line(4,-3){4}}
 \put(56,54){\framebox(3,3){}}
\put(52,62){\line(1,0){8}}       \put(56,63){\pos{bc}{$X_3$}}
\put(60,60){\framebox(4,4){}}
\put(64,62){\line(1,0){8}}
\put(72,60){\framebox(4,4){$=$}}
 \put(76,60){\line(4,-3){4}}
 \put(80,54){\framebox(3,3){}}
\put(2,54){\line(0,1){6}}
\put(0,50){\framebox(4,4){}}
\put(2,50){\line(0,-1){6}}
\put(26,54){\line(0,1){6}}
\put(24,50){\framebox(4,4){}}
\put(26,50){\line(0,-1){6}}
\put(50,54){\line(0,1){6}}
\put(48,50){\framebox(4,4){}}
\put(50,50){\line(0,-1){6}}
\put(74,54){\line(0,1){6}}
\put(72,50){\framebox(4,4){}}
\put(74,50){\line(0,-1){6}}
\put(0,40){\framebox(4,4){$=$}}
 \put(4,40){\line(4,-3){4}}
 \put(8,34){\framebox(3,3){}}
\put(4,42){\line(1,0){8}}
\put(12,40){\framebox(4,4){}}
\put(16,42){\line(1,0){8}}
\put(24,40){\framebox(4,4){$=$}}
 \put(28,40){\line(4,-3){4}}
 \put(32,34){\framebox(3,3){}}
\put(28,42){\line(1,0){8}}
\put(36,40){\framebox(4,4){}}
\put(40,42){\line(1,0){8}}
\put(48,40){\framebox(4,4){$=$}}
 \put(52,40){\line(4,-3){4}}
 \put(56,34){\framebox(3,3){}}
\put(52,42){\line(1,0){8}}
\put(60,40){\framebox(4,4){}}
\put(64,42){\line(1,0){8}}
\put(72,40){\framebox(4,4){$=$}}
 \put(76,40){\line(4,-3){4}}
 \put(80,34){\framebox(3,3){}}
\put(2,34){\line(0,1){6}}
\put(0,30){\framebox(4,4){}}
\put(2,30){\line(0,-1){6}}
\put(26,34){\line(0,1){6}}
\put(24,30){\framebox(4,4){}}
\put(26,30){\line(0,-1){6}}
\put(50,34){\line(0,1){6}}
\put(48,30){\framebox(4,4){}}
\put(50,30){\line(0,-1){6}}
\put(74,34){\line(0,1){6}}
\put(72,30){\framebox(4,4){}}
\put(74,30){\line(0,-1){6}}
\put(0,20){\framebox(4,4){$=$}}
 \put(4,20){\line(4,-3){4}}
 \put(8,14){\framebox(3,3){}}
\put(4,22){\line(1,0){8}}
\put(12,20){\framebox(4,4){}}
\put(16,22){\line(1,0){8}}
\put(24,20){\framebox(4,4){$=$}}
 \put(28,20){\line(4,-3){4}}
 \put(32,14){\framebox(3,3){}}
\put(28,22){\line(1,0){8}}
\put(36,20){\framebox(4,4){}}
\put(40,22){\line(1,0){8}}
\put(48,20){\framebox(4,4){$=$}}
 \put(52,20){\line(4,-3){4}}
 \put(56,14){\framebox(3,3){}}
\put(52,22){\line(1,0){8}}
\put(60,20){\framebox(4,4){}}
\put(64,22){\line(1,0){8}}
\put(72,20){\framebox(4,4){$=$}}
 \put(76,20){\line(4,-3){4}}
 \put(80,14){\framebox(3,3){}}
\put(2,14){\line(0,1){6}}
\put(0,10){\framebox(4,4){}}
\put(2,10){\line(0,-1){6}}
\put(26,14){\line(0,1){6}}
\put(24,10){\framebox(4,4){}}
\put(26,10){\line(0,-1){6}}
\put(50,14){\line(0,1){6}}
\put(48,10){\framebox(4,4){}}
\put(50,10){\line(0,-1){6}}
\put(74,14){\line(0,1){6}}
\put(72,10){\framebox(4,4){}}
\put(74,10){\line(0,-1){6}}
\put(0,0){\framebox(4,4){$=$}}
 \put(4,0){\line(4,-3){4}}
 \put(8,-6){\framebox(3,3){}}
\put(4,2){\line(1,0){8}}
\put(12,0){\framebox(4,4){}}
\put(16,2){\line(1,0){8}}
\put(24,0){\framebox(4,4){$=$}}
 \put(28,0){\line(4,-3){4}}
\put(32,-6){\framebox(3,3){}}
\put(28,2){\line(1,0){8}}
\put(36,0){\framebox(4,4){}}
\put(40,2){\line(1,0){8}}
\put(48,0){\framebox(4,4){$=$}}
 \put(52,0){\line(4,-3){4}}
 \put(56,-6){\framebox(3,3){}}
\put(52,2){\line(1,0){8}}
\put(60,0){\framebox(4,4){}}
\put(64,2){\line(1,0){8}}
\put(72,0){\framebox(4,4){$=$}}
 \put(76,0){\line(4,-3){4}}
 \put(80,-6){\framebox(3,3){}}
\end{picture}
%
\vspace{3ex}
\caption{\label{fig:2DGrid}%
Forney factor graph of the 2D Ising model in 
an external field, where
the unlabeled normal-size 
boxes represent factors as in~(\ref{eqn:IsingA}), the
small boxes represent factors as in~(\ref{eqn:IsingH}), and 
the boxes containing $=$ 
symbols are equality constraints.
}
\end{figure}
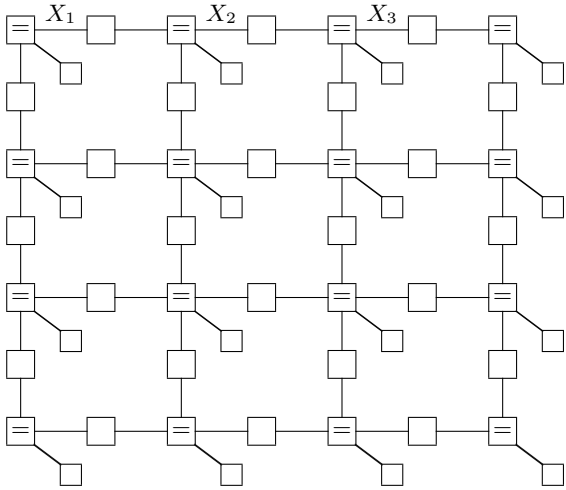

Let $X_1, X_2, \ldots, X_N$ be random variables
arranged on the sites of a 2D lattice,
as illustrated in Fig.~\ref{fig:2DGrid}, where interaction 
is restricted to 
adjacent (nearest-neighbor) variables. 
Suppose each random
variable takes values in a finite alphabet $\calX$.
Let $x_i$ represent a
possible realization of $X_i$, let $\x$ stand for 
a configuration $(x_1, x_2, \ldots, x_N)$, and let $\X$ stand for 
$(X_1, X_2, \ldots, X_N)$. 

We start with the 2D Ising model, generalizations to the $q$-state Potts
model are deferred to Section~\ref{sec:Potts}.
In a 2D Ising model, $\calX = \{0,1\}$ and the Hamiltonian 
is defined as~\cite{Yeo:92}
\begin{multline}
\label{eqn:HamiltonianI}
\mathcal{H}_{\text{Ising}}(\x) \eqdef -\!\!\sum_{\text{$(k,\ell)\in \B$}}\!\!\!J_{k, \ell}\cdot
\big([x_k = x_{\ell}] - [x_k \ne x_{\ell}]\big)\\ 
- \sum_{m = 1}^N H_m\cdot\big([x_m = 1] - [x_m = 0]\big)
\end{multline}
where the set $\B$ contains 
all the unordered pairs (bonds) $(k,\ell)$ with non-zero 
interactions and $[\cdot]$ denotes the Iverson 
bracket~\cite{Knuth:92}, which evaluates to $1$ if the condition in 
the bracket is satisfied and to $0$ otherwise.

The real coupling parameter $J_{k, \ell}$ controls the strength of
the interaction between adjacent variables $(x_k, x_{\ell})$.
The real parameter $H_m$ corresponds to the
presence of an external magnetic field and controls the strength of
the interaction between $X_m$ and the field.

In this paper, we concentrate on ferromagnetic models,
characterized by $J_{k, \ell} > 0$, for each $(k, \ell) \in \B$.
The external field is assumed to be consistent.
If $H_m > 0$, variable $X_m$ tends to have
value $1$, while $X_m$ tends to have
value $0$ if $H_m < 0$.

In thermal equilibrium, the probability that the model is in configuration $\x$ is
given by the Boltzmann distribution
\begin{equation} 
\label{eqn:Prob}
p_{\text{B}}(\x) \eqdef \frac{e^{-\beta \mathcal{H}(\x)}}{Z} 
\end{equation}

Here, the normalization constant $Z$ is the partition function $Z = \sum_{\x \in \calX^N} e^{-\beta \mathcal{H}(\x)}$ and
$\beta \eqdef \frac{1}{k_{\text{B}}T}$, where $T$ denotes the temperature 
and $k_{\text{B}}$ is Boltzmann's constant~\cite{Yeo:92, Cipra:87}. 

The Helmholtz free energy is defined as
\begin{equation} 
\label{eqn:FreeEnergy}
F_{\text{H}} \eqdef -\frac{1}{\beta}\ln Z
\end{equation}

In the rest of this paper, we will assume $\beta = 1$.
With this 
assumption, e.g., 
large values of $J$ and $|H|$ correspond to models at low temperature
and in a strong external 
field. 

For each adjacent pair $(x_k, x_\ell)$, let 
\begin{equation}
\label{eqn:IsingA}
\kappa_{k, \ell}(x_k, x_{\ell}) 
= e^{J_{k, \ell}\cdot\big([x_k = x_{\ell}] - [x_k \ne x_{\ell}]\big)}
\end{equation}
and for each $x_m$
\begin{equation}
\label{eqn:IsingH}
\tau_{m}(x_m) = e^{H_m\cdot\big([x_m = 1] - [x_m = 0]\big)}
\end{equation}

We can then define $f: \calX^N \rightarrow \R$ as 
\begin{equation} 
\label{eqn:factorF}
f(\x) \, \eqdef  \!\!\prod_{\text{$(k,\ell)\in \B$}}\!\!\!\kappa_{k, \ell}(x_k, x_{\ell})
 \prod_{m = 1}^N \tau_{m}(x_m)
\end{equation}

The corresponding Forney factor graph (normal graph) 
for the factorization
in~(\ref{eqn:factorF}) is shown in~\Fig{fig:2DGrid},
where the boxes labeled ``$=$'' are equality 
constraints~\cite{Lg:ifg2004}. 

From (\ref{eqn:factorF}), the partition function (\ref{eqn:Prob})
can be expressed as 
\begin{equation}
\label{eqn:PartFunction}
Z = \sum_{\x \in \calX^N} f(\x) 
\end{equation}

At high temperatures (i.e., small $J$), 
the Boltzmann distribution~(\ref{eqn:Prob})
approaches the uniform distribution. To estimate $Z$ in this case, Monte Carlo methods 
in the original factor graph, as in Fig.~\ref{fig:2DGrid}, generally perform well.
In this paper, we propose an importance
sampling scheme to compute an estimate of the partition function, where the only requirement to obtain
fast convergence is having a strong external
magnetic field (i.e., large $|H|$).
The scheme operates in 
the dual of the Forney factor graph representing the factorization 
in~(\ref{eqn:factorF}).

\begin{figure}[t]
\setlength{\unitlength}{0.89mm}
\centering
\begin{picture}(81,66)(0,0)
\small
\put(0,60){\framebox(4,4){$+$}}
 \put(4,60){\line(4,-3){4}}
 \put(8,54){\framebox(3,3){}}
\put(4,62){\line(1,0){8}}        
\put(12,60){\framebox(4,4){}}
\put(16,62){\line(1,0){8}}
\put(24,60){\framebox(4,4){$+$}}
 \put(28,60){\line(4,-3){4}}
 \put(32,54){\framebox(3,3){}}
\put(28,62){\line(1,0){8}}       
\put(36,60){\framebox(4,4){}}
\put(40,62){\line(1,0){8}}
\put(48,60){\framebox(4,4){$+$}}
 \put(52,60){\line(4,-3){4}}
 \put(56,54){\framebox(3,3){}}
\put(52,62){\line(1,0){8}}       
\put(60,60){\framebox(4,4){}}
\put(64,62){\line(1,0){8}}
\put(72,60){\framebox(4,4){$+$}}
 \put(76,60){\line(4,-3){4}}
 \put(80,54){\framebox(3,3){}}
\put(2,54){\line(0,1){6}}
\put(0,50){\framebox(4,4){}}
\put(2,50){\line(0,-1){6}}
\put(26,54){\line(0,1){6}}
\put(24,50){\framebox(4,4){}}
\put(26,50){\line(0,-1){6}}
\put(50,54){\line(0,1){6}}
\put(48,50){\framebox(4,4){}}
\put(50,50){\line(0,-1){6}}
\put(74,54){\line(0,1){6}}
\put(72,50){\framebox(4,4){}}
\put(74,50){\line(0,-1){6}}
\put(0,40){\framebox(4,4){$+$}}
 \put(4,40){\line(4,-3){4}}
 \put(8,34){\framebox(3,3){}}
\put(4,42){\line(1,0){8}}
\put(12,40){\framebox(4,4){}}
\put(16,42){\line(1,0){8}}
\put(24,40){\framebox(4,4){$+$}}
 \put(28,40){\line(4,-3){4}}
 \put(32,34){\framebox(3,3){}}
\put(28,42){\line(1,0){8}}
\put(36,40){\framebox(4,4){}}
\put(40,42){\line(1,0){8}}
\put(48,40){\framebox(4,4){$+$}}
 \put(52,40){\line(4,-3){4}}
 \put(56,34){\framebox(3,3){}}
\put(52,42){\line(1,0){8}}
\put(60,40){\framebox(4,4){}}
\put(64,42){\line(1,0){8}}
\put(72,40){\framebox(4,4){$+$}}
 \put(76,40){\line(4,-3){4}}
 \put(80,34){\framebox(3,3){}}
\put(2,34){\line(0,1){6}}
\put(0,30){\framebox(4,4){}}
\put(2,30){\line(0,-1){6}}
\put(26,34){\line(0,1){6}}
\put(24,30){\framebox(4,4){}}
\put(26,30){\line(0,-1){6}}
\put(50,34){\line(0,1){6}}
\put(48,30){\framebox(4,4){}}
\put(50,30){\line(0,-1){6}}
\put(74,34){\line(0,1){6}}
\put(72,30){\framebox(4,4){}}
\put(74,30){\line(0,-1){6}}
\put(0,20){\framebox(4,4){$+$}}
 \put(4,20){\line(4,-3){4}}
 \put(8,14){\framebox(3,3){}}
\put(4,22){\line(1,0){8}}
\put(12,20){\framebox(4,4){}}
\put(16,22){\line(1,0){8}}
\put(24,20){\framebox(4,4){$+$}}
 \put(28,20){\line(4,-3){4}}
 \put(32,14){\framebox(3,3){}}
\put(28,22){\line(1,0){8}}
\put(36,20){\framebox(4,4){}}
\put(40,22){\line(1,0){8}}
\put(48,20){\framebox(4,4){$+$}}
 \put(52,20){\line(4,-3){4}}
 \put(56,14){\framebox(3,3){}}
\put(52,22){\line(1,0){8}}
\put(60,20){\framebox(4,4){}}
\put(64,22){\line(1,0){8}}
\put(72,20){\framebox(4,4){$+$}}
 \put(76,20){\line(4,-3){4}}
 \put(80,14){\framebox(3,3){}}
\put(2,14){\line(0,1){6}}
\put(0,10){\framebox(4,4){}}
\put(2,10){\line(0,-1){6}}
\put(26,14){\line(0,1){6}}
\put(24,10){\framebox(4,4){}}
\put(26,10){\line(0,-1){6}}
\put(50,14){\line(0,1){6}}
\put(48,10){\framebox(4,4){}}
\put(50,10){\line(0,-1){6}}
\put(74,14){\line(0,1){6}}
\put(72,10){\framebox(4,4){}}
\put(74,10){\line(0,-1){6}}
\put(0,0){\framebox(4,4){$+$}}
 \put(4,0){\line(4,-3){4}}
 \put(8,-6){\framebox(3,3){}}
\put(4,2){\line(1,0){8}}
\put(12,0){\framebox(4,4){}}
\put(16,2){\line(1,0){8}}
\put(24,0){\framebox(4,4){$+$}}
 \put(28,0){\line(4,-3){4}}
\put(32,-6){\framebox(3,3){}}
\put(28,2){\line(1,0){8}}
\put(36,0){\framebox(4,4){}}
\put(40,2){\line(1,0){8}}
\put(48,0){\framebox(4,4){$+$}}
 \put(52,0){\line(4,-3){4}}
 \put(56,-6){\framebox(3,3){}}
\put(52,2){\line(1,0){8}}
\put(60,0){\framebox(4,4){}}
\put(64,2){\line(1,0){8}}
\put(72,0){\framebox(4,4){$+$}}
 \put(76,0){\line(4,-3){4}}
 \put(80,-6){\framebox(3,3){}}
\put(8,63){\pos{bc}{$\tilde X_1$}}
\put(20,63){\pos{bc}{$\tilde X_2$}}
\put(32,63){\pos{bc}{$\tilde X_3$}}
\end{picture}
\vspace{3.0ex}
\caption{\label{fig:2DGridD}%
Dual Forney factor graph of the 2D Ising model in 
an external field, where  
the small boxes represent factors as in~(\ref{eqn:IsingKernelDual2}), the 
unlabeled normal-size boxes represent factors 
as in~(\ref{eqn:IsingKernelD}), and the boxes 
containing $+$ symbols represent XOR factors as in~(\ref{eqn:XOR}).}
\vspace*{-0.8mm}
\end{figure}
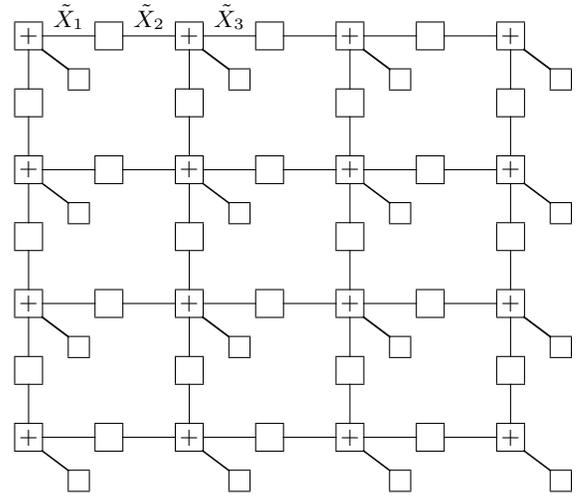

\section{The Dual Model}
\label{sec:NFGD}

Starting from a Forney factor graph, as in~\Fig{fig:2DGrid},
we can obtain its dual by 
replacing each variable $x$ with its dual
variable $\tilde x$, each 
factor $\kappa_{k, \ell}$ with its 2D Discrete Fourier 
transform (DFT)\footnote{Here, $\gamma(\tilde x_1, \tilde x_2)$, the 2D DFT of $\kappa(x_1, x_2)$, 
is defined as
\begin{equation*} 
\gamma(\tilde x_1, \tilde x_2) \eqdef 
\sum_{x_1\in \calX}\sum_{x_2 \in \calX} \kappa(x_1,x_2)
e^{-i2\pi(x_1\tilde x_1 + x_2\tilde x_2)/|\calX|}
\end{equation*}
where $i $ is the unit imaginary number~\cite{Brace:1999}.\vspace{0.5mm}
}, 
each factor $\tau_m$ with its one-dimensional (1D) 
DFT,
and each equality 
constraint with an XOR factor, see~\cite{Forney:01, Forney:11, AY:2011, FV:2011}.

In the dual domain, random variables $\tilde X$
also take their values in $\calX$.
The partition function is denoted
by $Z_\text{d}$ and the number of edges by $E$. 
For the models that we consider in this paper,
the normal factor graph duality theorem states 
that\footnote{To be more precise,
in our models there are no variables involved with only one 
factor. In Forney factor graphs such variables are represented by 
half-edges~\cite{Forney:01}. For
the general form of the normal factor graph 
duality theorem see~\cite{Forney:11,AY:2011}.}
\begin{equation}
\label{eqn:NDual}
Z_\text{d} = |\calX^{E}|Z
\end{equation}
see~\cite[Theorem 2]{AY:2011} and \cite{Forney:11}
especially for the normal factor graph duality theorem in the 
context of linear codes.


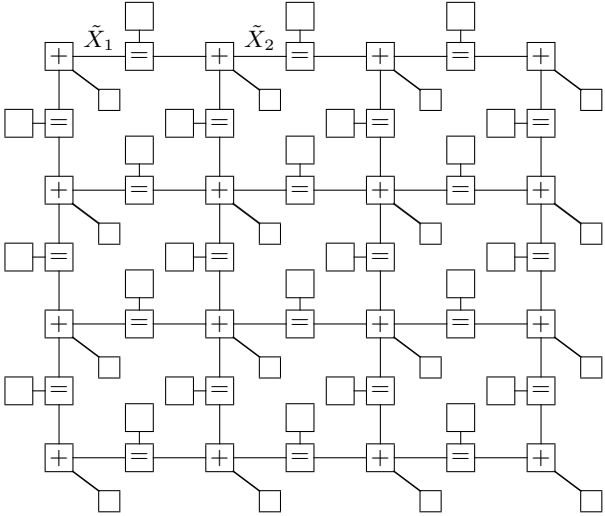
\begin{figure}[t]
\setlength{\unitlength}{0.89mm}
\centering
\begin{picture}(77,71.6)(0,0)
\small
\put(0,60){\framebox(4,4){$+$}}
 \put(4,60){\line(4,-3){4}}
 \put(8,54){\framebox(3,3){}}
\put(4,62){\line(1,0){8}}        
\put(12,60){\framebox(4,4){$=$}}
\put(16,62){\line(1,0){8}}
\put(24,60){\framebox(4,4){$+$}}
 \put(28,60){\line(4,-3){4}}
 \put(32,54){\framebox(3,3){}}
\put(28,62){\line(1,0){8}}       
\put(36,60){\framebox(4,4){$=$}}
\put(40,62){\line(1,0){8}}
\put(48,60){\framebox(4,4){$+$}}
 \put(52,60){\line(4,-3){4}}
 \put(56,54){\framebox(3,3){}}
\put(52,62){\line(1,0){8}}       
\put(60,60){\framebox(4,4){$=$}}
\put(64,62){\line(1,0){8}}
\put(72,60){\framebox(4,4){$+$}}
 \put(76,60){\line(4,-3){4}}
 \put(80,54){\framebox(3,3){}}
\put(2,54){\line(0,1){6}}
\put(0,50){\framebox(4,4){$=$}}
\put(2,50){\line(0,-1){6}}
\put(26,54){\line(0,1){6}}
\put(24,50){\framebox(4,4){$=$}}
\put(26,50){\line(0,-1){6}}
\put(50,54){\line(0,1){6}}
\put(48,50){\framebox(4,4){$=$}}
\put(50,50){\line(0,-1){6}}
\put(74,54){\line(0,1){6}}
\put(72,50){\framebox(4,4){$=$}}
\put(74,50){\line(0,-1){6}}
\put(0,40){\framebox(4,4){$+$}}
 \put(4,40){\line(4,-3){4}}
 \put(8,34){\framebox(3,3){}}
\put(4,42){\line(1,0){8}}
\put(12,40){\framebox(4,4){$=$}}
\put(16,42){\line(1,0){8}}
\put(24,40){\framebox(4,4){$+$}}
 \put(28,40){\line(4,-3){4}}
 \put(32,34){\framebox(3,3){}}
\put(28,42){\line(1,0){8}}
\put(36,40){\framebox(4,4){$=$}}
\put(40,42){\line(1,0){8}}
\put(48,40){\framebox(4,4){$+$}}
 \put(52,40){\line(4,-3){4}}
 \put(56,34){\framebox(3,3){}}
\put(52,42){\line(1,0){8}}
\put(60,40){\framebox(4,4){$=$}}
\put(64,42){\line(1,0){8}}
\put(72,40){\framebox(4,4){$+$}}
 \put(76,40){\line(4,-3){4}}
 \put(80,34){\framebox(3,3){}}
\put(2,34){\line(0,1){6}}
\put(0,30){\framebox(4,4){$=$}}
\put(2,30){\line(0,-1){6}}
\put(26,34){\line(0,1){6}}
\put(24,30){\framebox(4,4){$=$}}
\put(26,30){\line(0,-1){6}}
\put(50,34){\line(0,1){6}}
\put(48,30){\framebox(4,4){$=$}}
\put(50,30){\line(0,-1){6}}
\put(74,34){\line(0,1){6}}
\put(72,30){\framebox(4,4){$=$}}
\put(74,30){\line(0,-1){6}}
\put(0,20){\framebox(4,4){$+$}}
 \put(4,20){\line(4,-3){4}}
 \put(8,14){\framebox(3,3){}}
\put(4,22){\line(1,0){8}}
\put(12,20){\framebox(4,4){$=$}}
\put(16,22){\line(1,0){8}}
\put(24,20){\framebox(4,4){$+$}}
 \put(28,20){\line(4,-3){4}}
 \put(32,14){\framebox(3,3){}}
\put(28,22){\line(1,0){8}}
\put(36,20){\framebox(4,4){$=$}}
\put(40,22){\line(1,0){8}}
\put(48,20){\framebox(4,4){$+$}}
 \put(52,20){\line(4,-3){4}}
 \put(56,14){\framebox(3,3){}}
\put(52,22){\line(1,0){8}}
\put(60,20){\framebox(4,4){$=$}}
\put(64,22){\line(1,0){8}}
\put(72,20){\framebox(4,4){$+$}}
 \put(76,20){\line(4,-3){4}}
 \put(80,14){\framebox(3,3){}}
\put(2,14){\line(0,1){6}}
\put(0,10){\framebox(4,4){$=$}}
\put(2,10){\line(0,-1){6}}
\put(26,14){\line(0,1){6}}
\put(24,10){\framebox(4,4){$=$}}
\put(26,10){\line(0,-1){6}}
\put(50,14){\line(0,1){6}}
\put(48,10){\framebox(4,4){$=$}}
\put(50,10){\line(0,-1){6}}
\put(74,14){\line(0,1){6}}
\put(72,10){\framebox(4,4){$=$}}
\put(74,10){\line(0,-1){6}}
\put(0,0){\framebox(4,4){$+$}}
 \put(4,0){\line(4,-3){4}}
 \put(8,-6){\framebox(3,3){}}
\put(4,2){\line(1,0){8}}
\put(12,0){\framebox(4,4){$=$}}
\put(16,2){\line(1,0){8}}
\put(24,0){\framebox(4,4){$+$}}
 \put(28,0){\line(4,-3){4}}
\put(32,-6){\framebox(3,3){}}
\put(28,2){\line(1,0){8}}
\put(36,0){\framebox(4,4){$=$}}
\put(40,2){\line(1,0){8}}
\put(48,0){\framebox(4,4){$+$}}
 \put(52,0){\line(4,-3){4}}
 \put(56,-6){\framebox(3,3){}}
\put(52,2){\line(1,0){8}}
\put(60,0){\framebox(4,4){$=$}}
\put(64,2){\line(1,0){8}}
\put(72,0){\framebox(4,4){$+$}}
 \put(76,0){\line(4,-3){4}}
 \put(80,-6){\framebox(3,3){}}
 
\put(8,63){\pos{bc}{$\tilde X_1$}}
\put(32,63){\pos{bc}{$\tilde X_2$}}

\put(14,64){\line(0,1){2}}
\put(38,64){\line(0,1){2}}
\put(62,64){\line(0,1){2}}
\put(12,66){\framebox(4,4){$$}}
\put(36,66){\framebox(4,4){$$}}
\put(60,66){\framebox(4,4){$$}}
\put(14,44){\line(0,1){2}}
\put(38,44){\line(0,1){2}}
\put(62,44){\line(0,1){2}}
\put(12,46){\framebox(4,4){$$}}
\put(36,46){\framebox(4,4){$$}}
\put(60,46){\framebox(4,4){$$}}
\put(14,24){\line(0,1){2}}
\put(38,24){\line(0,1){2}}
\put(62,24){\line(0,1){2}}
\put(12,26){\framebox(4,4){$$}}
\put(36,26){\framebox(4,4){$$}}
\put(60,26){\framebox(4,4){$$}}
\put(14,4){\line(0,1){2}}
\put(38,4){\line(0,1){2}}
\put(62,4){\line(0,1){2}}
\put(12,6){\framebox(4,4){$$}}
\put(36,6){\framebox(4,4){$$}}
\put(60,6){\framebox(4,4){$$}}
\put(0,52){\line(-1,0){2}}
\put(24,52){\line(-1,0){2}}
\put(48,52){\line(-1,0){2}}
\put(72,52){\line(-1,0){2}}
\put(-6,50){\framebox(4,4){$$}}
\put(18,50){\framebox(4,4){$$}}
\put(42,50){\framebox(4,4){$$}}
\put(66,50){\framebox(4,4){$$}}
\put(0,32){\line(-1,0){2}}
\put(24,32){\line(-1,0){2}}
\put(48,32){\line(-1,0){2}}
\put(72,32){\line(-1,0){2}}
\put(-6,30){\framebox(4,4){$$}}
\put(18,30){\framebox(4,4){$$}}
\put(42,30){\framebox(4,4){$$}}
\put(66,30){\framebox(4,4){$$}}
\put(0,12){\line(-1,0){2}}
\put(24,12){\line(-1,0){2}}
\put(48,12){\line(-1,0){2}}
\put(72,12){\line(-1,0){2}}
\put(-6,10){\framebox(4,4){$$}}
\put(18,10){\framebox(4,4){$$}}
\put(42,10){\framebox(4,4){$$}}
\put(66,10){\framebox(4,4){$$}}
\end{picture}
\vspace{3.0ex}
\caption{\label{fig:2DGridDM}%
Modified dual Forney factor graph of the 2D Ising model in an external 
field, where the small 
boxes represent factors as in~(\ref{eqn:IsingKernelDual2}), the unlabeled 
normal-size boxes 
represent factors as in~(\ref{eqn:IsingKernelDual}), 
and boxes containing 
$+$ symbols represent XOR factors as in~(\ref{eqn:XOR}). 
\vspace{0.5mm}
}
\end{figure}


For variables $\tilde x_1, \tilde x_2, \ldots, \tilde x_k$, in the dual Forney factor graph of the 
Ising model, XOR factors are defined as
\begin{equation} 
\label{eqn:XOR}
g(\tilde x_1, \tilde x_2, \ldots, \tilde x_k) \eqdef 
[\tilde x_1 \oplus \tilde x_2 \oplus \ldots \oplus \tilde x_k=0]
\end{equation}
where $\oplus$ denotes addition in GF($2$).

The 1D DFT of factors as in~(\ref{eqn:IsingH}) will have the
following form
\begin{equation} 
\label{eqn:IsingKernelDual2}
\lambda_{m}(\tilde x_m) = \left\{ \begin{array}{lr}
     2\cosh H_{m}, & \text{if $\tilde x_m = 0$} \\
     -2\sinh H_{m}, & \text{if $\tilde x_m = 1$} 
\end{array} \right.
\end{equation}
and each factor~(\ref{eqn:IsingA}) is replaced by its 2D DFT,
which has the following form 
\begin{equation} 
\label{eqn:IsingKernelD}
\gamma_{k, \ell}(\tilde x_k, \tilde x_\ell) = \left\{ \begin{array}{ll}
     4\cosh J_{k, \ell}, & \text{if $\tilde x_k = \tilde x_\ell = 0$} \\
     4\sinh J_{k, \ell}, & \text{if $\tilde x_k = \tilde x_\ell = 1$} \\
     0, & \text{otherwise.}
  \end{array} \right.
\end{equation}

The corresponding dual Forney factor graph with factors as 
in~(\ref{eqn:IsingKernelDual2}) and~(\ref{eqn:IsingKernelD})  
is shown in~\Fig{fig:2DGridD}.

All the factors in~(\ref{eqn:IsingKernelD}) are diagonal, therefore
it is possible to 
simplify the dual factor graph in~\Fig{fig:2DGridD},
to construct the modified dual factor graph depicted in~\Fig{fig:2DGridDM},
with factors attached to each equality constraint as 
\begin{equation} 
\label{eqn:IsingKernelDual}
\gamma_{k}(\tilde x_k) = \left\{ \begin{array}{ll}
     4\cosh J_{k}, & \text{if $\tilde x_k = 0$} \\
     4\sinh J_{k}, & \text{if $\tilde x_k = 1$}
  \end{array} \right.
\end{equation}

Here, $J_k$ is 
the coupling parameter associated with each bond (the bond strength). 
The corresponding modified dual Forney factor graph with factors 
as in~(\ref{eqn:IsingKernelDual2}) and~(\ref{eqn:IsingKernelDual}) 
is shown in~\Fig{fig:2DGridDM}; see~\cite{MoLo:ISIT2013} for more details.

In this paper, we 
concentrate
on ferromagnetic models, therefore all
the factors as in~(\ref{eqn:IsingKernelDual}) are positive.
Since in a 2D Ising model, the value of $Z$ is invariant under the
change of sign of the external magnetic field~\cite{Baxter07}, without
loss of generality,  
we assume $H_m <  0$. With this assumption, all the factors 
as in~(\ref{eqn:IsingKernelDual2}) will also be 
positive\footnote{The factors in the dual domain can in general be negative or 
complex-valued. Here, we require all factors to be positive
because we need to define a probability 
mass function on the dual factor graph, which is then used in the importance sampling scheme of 
Section~\ref{sec:IS}.
Applying Monte Carlo methods to factor graphs with negative and complex 
factors is discussed in~\cite{MoLo:ITW2012}.
}.


\begin{figure}[t]
\setlength{\unitlength}{0.89mm}
\centering
\begin{picture}(77,71.55)(0,0)
\small
\put(0,60){\framebox(4,4){$=$}}
 \put(4,60){\line(4,-3){4}}
 \put(8,54){\framebox(3,3){}}
\put(4,62){\line(1,0){8}}        
\put(12,60){\framebox(4,4){$+$}}
\put(16,62){\line(1,0){8}}
\put(24,60){\framebox(4,4){$=$}}
 \put(28,60){\line(4,-3){4}}
 \put(32,54){\framebox(3,3){}}
\put(28,62){\line(1,0){8}}       
\put(36,60){\framebox(4,4){$+$}}
\put(40,62){\line(1,0){8}}
\put(48,60){\framebox(4,4){$=$}}
 \put(52,60){\line(4,-3){4}}
 \put(56,54){\framebox(3,3){}}
\put(52,62){\line(1,0){8}}       
\put(60,60){\framebox(4,4){$+$}}
\put(64,62){\line(1,0){8}}
\put(72,60){\framebox(4,4){$=$}}
 \put(76,60){\line(4,-3){4}}
 \put(80,54){\framebox(3,3){}}
\put(2,54){\line(0,1){6}}
\put(0,50){\framebox(4,4){$+$}}
\put(2,50){\line(0,-1){6}}
\put(26,54){\line(0,1){6}}
\put(24,50){\framebox(4,4){$+$}}
\put(26,50){\line(0,-1){6}}
\put(50,54){\line(0,1){6}}
\put(48,50){\framebox(4,4){$+$}}
\put(50,50){\line(0,-1){6}}
\put(74,54){\line(0,1){6}}
\put(72,50){\framebox(4,4){$+$}}
\put(74,50){\line(0,-1){6}}
\put(0,40){\framebox(4,4){$=$}}
 \put(4,40){\line(4,-3){4}}
 \put(8,34){\framebox(3,3){}}
\put(4,42){\line(1,0){8}}
\put(12,40){\framebox(4,4){$+$}}
\put(16,42){\line(1,0){8}}
\put(24,40){\framebox(4,4){$=$}}
 \put(28,40){\line(4,-3){4}}
 \put(32,34){\framebox(3,3){}}
\put(28,42){\line(1,0){8}}
\put(36,40){\framebox(4,4){$+$}}
\put(40,42){\line(1,0){8}}
\put(48,40){\framebox(4,4){$=$}}
 \put(52,40){\line(4,-3){4}}
 \put(56,34){\framebox(3,3){}}
\put(52,42){\line(1,0){8}}
\put(60,40){\framebox(4,4){$+$}}
\put(64,42){\line(1,0){8}}
\put(72,40){\framebox(4,4){$=$}}
 \put(76,40){\line(4,-3){4}}
 \put(80,34){\framebox(3,3){}}
\put(2,34){\line(0,1){6}}
\put(0,30){\framebox(4,4){$+$}}
\put(2,30){\line(0,-1){6}}
\put(26,34){\line(0,1){6}}
\put(24,30){\framebox(4,4){$+$}}
\put(26,30){\line(0,-1){6}}
\put(50,34){\line(0,1){6}}
\put(48,30){\framebox(4,4){$+$}}
\put(50,30){\line(0,-1){6}}
\put(74,34){\line(0,1){6}}
\put(72,30){\framebox(4,4){$+$}}
\put(74,30){\line(0,-1){6}}
\put(0,20){\framebox(4,4){$=$}}
 \put(4,20){\line(4,-3){4}}
 \put(8,14){\framebox(3,3){}}
\put(4,22){\line(1,0){8}}
\put(12,20){\framebox(4,4){$+$}}
\put(16,22){\line(1,0){8}}
\put(24,20){\framebox(4,4){$=$}}
 \put(28,20){\line(4,-3){4}}
 \put(32,14){\framebox(3,3){}}
\put(28,22){\line(1,0){8}}
\put(36,20){\framebox(4,4){$+$}}
\put(40,22){\line(1,0){8}}
\put(48,20){\framebox(4,4){$=$}}
 \put(52,20){\line(4,-3){4}}
 \put(56,14){\framebox(3,3){}}
\put(52,22){\line(1,0){8}}
\put(60,20){\framebox(4,4){$+$}}
\put(64,22){\line(1,0){8}}
\put(72,20){\framebox(4,4){$=$}}
 \put(76,20){\line(4,-3){4}}
 \put(80,14){\framebox(3,3){}}
\put(2,14){\line(0,1){6}}
\put(0,10){\framebox(4,4){$+$}}
\put(2,10){\line(0,-1){6}}
\put(26,14){\line(0,1){6}}
\put(24,10){\framebox(4,4){$+$}}
\put(26,10){\line(0,-1){6}}
\put(50,14){\line(0,1){6}}
\put(48,10){\framebox(4,4){$+$}}
\put(50,10){\line(0,-1){6}}
\put(74,14){\line(0,1){6}}
\put(72,10){\framebox(4,4){$+$}}
\put(74,10){\line(0,-1){6}}
\put(0,0){\framebox(4,4){$=$}}
 \put(4,0){\line(4,-3){4}}
 \put(8,-6){\framebox(3,3){}}
\put(4,2){\line(1,0){8}}
\put(12,0){\framebox(4,4){$+$}}
\put(16,2){\line(1,0){8}}
\put(24,0){\framebox(4,4){$=$}}
 \put(28,0){\line(4,-3){4}}
\put(32,-6){\framebox(3,3){}}
\put(28,2){\line(1,0){8}}
\put(36,0){\framebox(4,4){$+$}}
\put(40,2){\line(1,0){8}}
\put(48,0){\framebox(4,4){$=$}}
 \put(52,0){\line(4,-3){4}}
 \put(56,-6){\framebox(3,3){}}
\put(52,2){\line(1,0){8}}
\put(60,0){\framebox(4,4){$+$}}
\put(64,2){\line(1,0){8}}
\put(72,0){\framebox(4,4){$=$}}
 \put(76,0){\line(4,-3){4}}
 \put(80,-6){\framebox(3,3){}}
 
\put(8,63){\pos{bc}{$X_1$}}
\put(32,63){\pos{bc}{$X_2$}}

\put(14,64){\line(0,1){2}}
\put(38,64){\line(0,1){2}}
\put(62,64){\line(0,1){2}}
\put(12,66){\framebox(4,4){$$}}
\put(36,66){\framebox(4,4){$$}}
\put(60,66){\framebox(4,4){$$}}
\put(14,44){\line(0,1){2}}
\put(38,44){\line(0,1){2}}
\put(62,44){\line(0,1){2}}
\put(12,46){\framebox(4,4){$$}}
\put(36,46){\framebox(4,4){$$}}
\put(60,46){\framebox(4,4){$$}}
\put(14,24){\line(0,1){2}}
\put(38,24){\line(0,1){2}}
\put(62,24){\line(0,1){2}}
\put(12,26){\framebox(4,4){$$}}
\put(36,26){\framebox(4,4){$$}}
\put(60,26){\framebox(4,4){$$}}
\put(14,4){\line(0,1){2}}
\put(38,4){\line(0,1){2}}
\put(62,4){\line(0,1){2}}
\put(12,6){\framebox(4,4){$$}}
\put(36,6){\framebox(4,4){$$}}
\put(60,6){\framebox(4,4){$$}}
\put(0,52){\line(-1,0){2}}
\put(24,52){\line(-1,0){2}}
\put(48,52){\line(-1,0){2}}
\put(72,52){\line(-1,0){2}}
\put(-6,50){\framebox(4,4){$$}}
\put(18,50){\framebox(4,4){$$}}
\put(42,50){\framebox(4,4){$$}}
\put(66,50){\framebox(4,4){$$}}
\put(0,32){\line(-1,0){2}}
\put(24,32){\line(-1,0){2}}
\put(48,32){\line(-1,0){2}}
\put(72,32){\line(-1,0){2}}
\put(-6,30){\framebox(4,4){$$}}
\put(18,30){\framebox(4,4){$$}}
\put(42,30){\framebox(4,4){$$}}
\put(66,30){\framebox(4,4){$$}}
\put(0,12){\line(-1,0){2}}
\put(24,12){\line(-1,0){2}}
\put(48,12){\line(-1,0){2}}
\put(72,12){\line(-1,0){2}}
\put(-6,10){\framebox(4,4){$$}}
\put(18,10){\framebox(4,4){$$}}
\put(42,10){\framebox(4,4){$$}}
\put(66,10){\framebox(4,4){$$}}
\end{picture}
\vspace{3.0ex}
\caption{\label{fig:2DGridOrig}%
Modified Forney factor graph of the 2D Ising model in an external 
field, where the small 
boxes represent factors as in~(\ref{eqn:IsingKernelOrig2}), the unlabeled 
normal-size boxes 
represent factors as in~(\ref{eqn:IsingKernelOrig}), 
and boxes containing 
$+$ symbols represent XOR factors as in~(\ref{eqn:XOR}). 
\vspace*{0.7mm}
}
\end{figure}

As a side remark, we point out that by looking at the dual of the
modified dual Forney factor graph in~\Fig{fig:2DGridDM} we can obtain the
modified Forney factor graph of a 2D Ising model in an external 
field, with factors attached to each equality constraint as
\begin{equation} 
\label{eqn:IsingKernelOrig2}
\tau_{m}(x_m) = \left\{ \begin{array}{lr}
    e^{-H_m}, & \text{if $x_m = 0$} \\
     e^{H_m}, & \text{if $x_m = 1$} 
\end{array} \right.
\end{equation}
and with factors attached to each XOR
factor as
\begin{equation} 
\label{eqn:IsingKernelOrig}
\kappa_{k}(x_k) = \left\{ \begin{array}{ll}
     e^{J_k}, & \text{if $x_k = 0$} \\
     e^{-J_k}, & \text{if $x_k = 1$}
  \end{array} \right.
\end{equation}
The corresponding modified Forney factor graph is shown in~\Fig{fig:2DGridOrig}.

In Section~\ref{sec:IS}, we use the dual Forney factor graph representation of 
the Ising model to propose an importance sampling
scheme to compute an estimate of $Z$,
as in (\ref{eqn:PartFunction}).

\section{An Importance Sampling Scheme on Dual Factor Graphs}
\label{sec:IS}

We describe our importance sampling scheme on the modified dual 
Forney factor graph of the 2D Ising model
in an external 
field shown in~\Fig{fig:2DGridDM}. The scheme can be 
described analogously for
the 2D $q$-state Potts model, see Section~\ref{sec:Potts}.

Let us partition the set of random variables 
$\tilde \X$, into $\tilde \X_A$ and $\tilde \X_B$, with the restriction that the random 
variables in $\tilde \X_B$ are 
linear combinations (involving the XOR factors) of the random variables in
$\tilde \X_A$. 

\begin{figure}[t]
\setlength{\unitlength}{0.91mm}
\centering
\begin{picture}(77,72.25)(0,0)
\small
\put(0,60){\framebox(4,4){$+$}}
 \put(8,54){\framebox(3,3){}}
\put(12,60){\framebox(4,4){$=$}}
\put(24,60){\framebox(4,4){$+$}}
 \put(32,54){\framebox(3,3){}}
\put(36,60){\framebox(4,4){$=$}}
\put(48,60){\framebox(4,4){$+$}}
 \put(56,54){\framebox(3,3){}}
\put(60,60){\framebox(4,4){$=$}}
\put(72,60){\framebox(4,4){$+$}}
 \put(80,54){\framebox(3,3){}}
 \put(0,50){\framebox(4,4){$=$}}
\put(24,50){\framebox(4,4){$=$}}
\put(48,50){\framebox(4,4){$=$}}
\put(72,50){\framebox(4,4){$=$}}
\put(0,40){\framebox(4,4){$+$}}
 \put(8,34){\framebox(3,3){}}
\put(12,40){\framebox(4,4){$=$}}
\put(24,40){\framebox(4,4){$+$}}
 \put(32,34){\framebox(3,3){}}
\put(36,40){\framebox(4,4){$=$}}
\put(48,40){\framebox(4,4){$+$}}
 \put(56,34){\framebox(3,3){}}
\put(60,40){\framebox(4,4){$=$}}
\put(72,40){\framebox(4,4){$+$}}
 \put(80,34){\framebox(3,3){}}
\put(0,30){\framebox(4,4){$=$}}
\put(24,30){\framebox(4,4){$=$}}
\put(48,30){\framebox(4,4){$=$}}
\put(72,30){\framebox(4,4){$=$}}
\put(0,20){\framebox(4,4){$+$}}
 \put(8,14){\framebox(3,3){}}
\put(12,20){\framebox(4,4){$=$}}
\put(24,20){\framebox(4,4){$+$}}
 \put(32,14){\framebox(3,3){}}
\put(36,20){\framebox(4,4){$=$}}
\put(48,20){\framebox(4,4){$+$}}
 \put(56,14){\framebox(3,3){}}
\put(60,20){\framebox(4,4){$=$}}
\put(72,20){\framebox(4,4){$+$}}
 \put(80,14){\framebox(3,3){}}
\put(0,10){\framebox(4,4){$=$}}
\put(24,10){\framebox(4,4){$=$}}
\put(48,10){\framebox(4,4){$=$}}
\put(72,10){\framebox(4,4){$=$}}
\put(0,0){\framebox(4,4){$+$}}
 \put(8,-6){\framebox(3,3){}}
\put(12,0){\framebox(4,4){$=$}}
\put(24,0){\framebox(4,4){$+$}}
\put(32,-6){\framebox(3,3){}}
\put(36,0){\framebox(4,4){$=$}}
\put(48,0){\framebox(4,4){$+$}}
 \put(56,-6){\framebox(3,3){}}
\put(60,0){\framebox(4,4){$=$}}
\put(72,0){\framebox(4,4){$+$}}
 \put(80,-6){\framebox(3,3){}}

\put(12,66){\framebox(4,4){$$}}
\put(36,66){\framebox(4,4){$$}}
\put(60,66){\framebox(4,4){$$}}
\put(12,46){\framebox(4,4){$$}}
\put(36,46){\framebox(4,4){$$}}
\put(60,46){\framebox(4,4){$$}}
\put(12,26){\framebox(4,4){$$}}
\put(36,26){\framebox(4,4){$$}}
\put(60,26){\framebox(4,4){$$}}
\put(12,6){\framebox(4,4){$$}}
\put(36,6){\framebox(4,4){$$}}
\put(60,6){\framebox(4,4){$$}}
\put(-6,50){\framebox(4,4){$$}}
\put(18,50){\framebox(4,4){$$}}
\put(42,50){\framebox(4,4){$$}}
\put(66,50){\framebox(4,4){$$}}
\put(-6,30){\framebox(4,4){$$}}
\put(18,30){\framebox(4,4){$$}}
\put(42,30){\framebox(4,4){$$}}
\put(66,30){\framebox(4,4){$$}}
\put(-6,10){\framebox(4,4){$$}}
\put(18,10){\framebox(4,4){$$}}
\put(42,10){\framebox(4,4){$$}}
\put(66,10){\framebox(4,4){$$}}

 \put(4,60){\line(4,-3){4}}
 \put(28,60){\line(4,-3){4}}
 \put(52,60){\line(4,-3){4}}
 \put(76,60){\line(4,-3){4}}

 \put(4,40){\line(4,-3){4}}
 \put(28,40){\line(4,-3){4}}
 \put(52,40){\line(4,-3){4}}
 \put(76,40){\line(4,-3){4}}
 \put(4,20){\line(4,-3){4}}
 \put(28,20){\line(4,-3){4}}
 \put(52,20){\line(4,-3){4}}
 \put(76,20){\line(4,-3){4}}
 \put(4,0){\line(4,-3){4}}
 \put(28,0){\line(4,-3){4}}
 \put(52,0){\line(4,-3){4}}
 \put(76,0){\line(4,-3){4}}

 \linethickness{0.67mm}
\put(14,4){\line(0,1){2}}
\put(38,4){\line(0,1){2}}
\put(62,4){\line(0,1){2}} 

 \put(14,64){\line(0,1){2}}
\put(38,64){\line(0,1){2}}
\put(62,64){\line(0,1){2}}

\put(14,44){\line(0,1){2}}
\put(38,44){\line(0,1){2}}
\put(62,44){\line(0,1){2}}

 \put(14,24){\line(0,1){2}}
\put(38,24){\line(0,1){2}}
\put(62,24){\line(0,1){2}}

\put(0,52){\line(-1,0){2}}
\put(24,52){\line(-1,0){2}}
\put(48,52){\line(-1,0){2}}
\put(72,52){\line(-1,0){2}}

\put(0,32){\line(-1,0){2}}
\put(24,32){\line(-1,0){2}}
\put(48,32){\line(-1,0){2}}
\put(72,32){\line(-1,0){2}}

\put(0,12){\line(-1,0){2}}
\put(24,12){\line(-1,0){2}}
\put(48,12){\line(-1,0){2}}
\put(72,12){\line(-1,0){2}}

 
\put(4,62){\line(1,0){8}}        
\put(16,62){\line(1,0){8}}
\put(28,62){\line(1,0){8}}       
\put(40,62){\line(1,0){8}}
\put(52,62){\line(1,0){8}}       
\put(64,62){\line(1,0){8}}
\put(2,54){\line(0,1){6}}
\put(2,50){\line(0,-1){6}}
\put(26,54){\line(0,1){6}}
\put(26,50){\line(0,-1){6}}
\put(50,54){\line(0,1){6}}
\put(50,50){\line(0,-1){6}}
\put(74,54){\line(0,1){6}}
\put(74,50){\line(0,-1){6}}
\put(4,42){\line(1,0){8}}
\put(16,42){\line(1,0){8}}
\put(28,42){\line(1,0){8}}
\put(40,42){\line(1,0){8}}
\put(52,42){\line(1,0){8}}
\put(64,42){\line(1,0){8}}
\put(2,34){\line(0,1){6}}
\put(2,30){\line(0,-1){6}}
\put(26,34){\line(0,1){6}}
\put(26,30){\line(0,-1){6}}
\put(50,34){\line(0,1){6}}
\put(50,30){\line(0,-1){6}}
\put(74,34){\line(0,1){6}}
\put(74,30){\line(0,-1){6}}
\put(4,22){\line(1,0){8}}
\put(16,22){\line(1,0){8}}
\put(28,22){\line(1,0){8}}
\put(40,22){\line(1,0){8}}
\put(52,22){\line(1,0){8}}
\put(64,22){\line(1,0){8}}
\put(2,14){\line(0,1){6}}
\put(2,10){\line(0,-1){6}}
\put(26,14){\line(0,1){6}}
\put(26,10){\line(0,-1){6}}
\put(50,14){\line(0,1){6}}
\put(50,10){\line(0,-1){6}}
\put(74,14){\line(0,1){6}}
\put(74,10){\line(0,-1){6}}
\put(4,2){\line(1,0){8}}
\put(16,2){\line(1,0){8}}
\put(28,2){\line(1,0){8}}
\put(40,2){\line(1,0){8}}
\put(52,2){\line(1,0){8}}
\put(64,2){\line(1,0){8}}
 
%
\end{picture}
\vspace{2.65ex}
\caption{\label{fig:2DGridDPart}%
A partitioning of variables on the modified dual Forney factor graph of the 2D Ising 
model in an external field, where the thick edges (bonds) represent variables 
in $\tilde \X_A$ and
edges connected to the unlabeled 
small boxes represent variables in $\tilde \X_B$. Here,
variables in $\tilde \X_B$ are
linear combinations (involving XOR factors) of the variables in $\tilde \X_A$.
}
\end{figure}
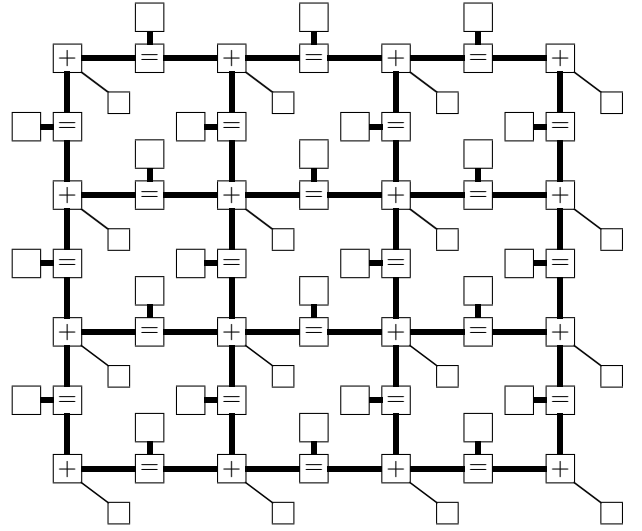

In our framework, we let $\tilde \X_B$ be the
set of all the variables represented by the edges connected to 
the small unlabeled boxes in Figs.~\ref{fig:2DGridDM} and~\ref{fig:2DGridDPart}, 
which
are the variables involved in factors as 
in~(\ref{eqn:IsingKernelDual2}). With this choice, $\tilde \X_A$ will 
contain all the variables involved in factors as in~(\ref{eqn:IsingKernelDual}), 
which are associated with each bond in the modified dual 
Forney factor graph and are marked by thick edges in~\Fig{fig:2DGridDPart}.
As will be discussed, this choice of partitioning is appropriate for models in a strong 
external field. A valid configuration $\tilde \x = (\tilde \x_A, \tilde \x_B)$ in the dual factor graph can
be created by assigning values to $\tilde \X_A$, followed by updating
$\tilde \X_B$ as linear combinations of $\tilde \X_A$.


Let us define
\begin{IEEEeqnarray}{r,C,l}
\Gamma(\tilde \x_A) & \eqdef & 
\prod_{\tilde x_k \in \tilde \x_A} \gamma _{k}(\tilde x_k) \label{eqn:PartG}\\
\Lambda(\tilde \x_B) & \eqdef & 
\prod_{\tilde x_m \in \tilde \x_B} \lambda_{m}(\tilde x_m)\label{eqn:PartL}
\end{IEEEeqnarray}

 We use the following probability mass function as the auxiliary distribution in our 
 importance sampling scheme
\begin{equation} 
\label{eqn:AuxDist}
q(\tilde \x_A) \eqdef \frac{\Gamma(\tilde \x_A)}{Z_q}, \, \qquad \forall \tilde \x_A \in \calX^{|\B|}
\end{equation}

The auxiliary distribution~(\ref{eqn:AuxDist})
has two key properties.
First, its partition function $Z_q$ is analytically available as
\begin{IEEEeqnarray}{r,C,l}
Z_q  & = & \sum_{\tilde \x_A} \Gamma(\tilde \x_A) \\
        & = & \prod_{k \in \B} 4(\cosh J_k + \sinh J_k) \\
        & = & 4^{|\B|} \text{exp}\Big(\sum_{k  \in \B} J_k\Big) \label{eqn:Zq} 
\end{IEEEeqnarray}
where $|\B|$ is the cardinality
of $\B$, which is equal to the number of bonds (the number of interacting pairs) in 
the lattice (cf.\ Section \ref{sec:Ising}). 
The value
of $Z_q$ is thus a function of the sum of all the
coupling parameters.


Second, it is straightforward to draw \emph{independent} samples 
$\tilde \x_A^{(1)}, \tilde \x_A^{(2)}, \ldots, \tilde \x_A^{(\ell)}, \ldots$,
according to $q(\tilde \x_A)$. To draw $\tilde \x_A^{(\ell)}$, 
we use the following algorithm 
\vspace{0.5ex}

\begin{itemize}
\itemsep1.8pt
\item[] \emph{draw} $u_1^{(\ell)}, u_2^{(\ell)}, \ldots, u_{|\B|}^{(\ell)}\overset{\text{i.i.d.}}{\sim} \, \mathcal{U}[0,1]$
\item[] {\bf for} $k = 1$ {\bf to} $|\B|$
\item [] \hspace{4.5mm} {\bf if} $u_k^{(\ell)} < \frac{1}{2}(1+e^{-2J_k})$
\item [] \hspace{10mm}  $\tilde x_{A,k}^{(\ell)} = 0$
\item [] \hspace{4.5mm} {\bf else}
\item [] \hspace{10mm}  $\tilde x_{A,k}^{(\ell)} = 1$
\item [] \hspace{4.5mm} {\bf end if}
\item[] {\bf end for}
\end{itemize}
\vspace{0.5ex}
The quantity
$\frac{1}{2}(1+e^{-2J_k})$ is equal to 
$\gamma_k(0)/\big(\gamma_k(0) + \gamma_k(1)\big)$.

\vspace{0.55ex}

Random variables in $\tilde \X_B$ are 
linear combinations 
of those in $\tilde \X_A$,
therefore after drawing $\tilde \x_A^{(\ell)}$, 
updating $\tilde \x_B^{(\ell)}$ can be done in a
straightforward manner. The samples are then used in the following 
importance sampling algorithm to estimate $Z_\text{d}/Z_{q}$.
\vspace{0.5ex}
\begin{itemize}
\itemsep1.8pt
\item[] \mbox{\emph{draw} $\tilde \x_A^{(1)}, \tilde \x_A^{(2)}, \ldots, \tilde \x_A^{(L)}$ \emph{according to} 
$q(\tilde \x_A)$}
\item[] \emph{update} $\tilde \x_B^{(1)}, \tilde \x_B^{(2)}, \ldots, \tilde \x_B^{(L)}$
\item[] \emph{compute}
\begin{equation} 
\label{eqn:EstR}
\hat r_{\text{IS}} = 
\frac{1}{L} \sum_{\ell = 1}^L \Lambda(\tilde \x_B^{(\ell)})
\end{equation}
\end{itemize}
\vspace{0.5ex}

It follows that, $\hat r_{\text{IS}}$ is an unbiased and consistent estimator 
of $Z_\text{d}/Z_{q}$. 

Indeed
\begin{equation}
\E_q[\, \hat r_{\text{IS}}\,] = \frac{Z_\text{d}}{Z_q}
\end{equation}


Since $Z_q$ is analytically available~(\ref{eqn:Zq}), the proposed
importance sampling scheme 
can yield an estimate of $Z_\text{d}$, which can then be used to estimate
the partition function~(\ref{eqn:PartFunction}), using the normal factor 
graph duality theorem~(cf.\ Section~\ref{sec:NFGD}).

The accuracy of $\hat r_{\text{IS}}$ in~(\ref{eqn:EstR}) depends on the
fluctuations of $\Lambda(\tilde \x_B)$. If $\Lambda(\tilde \x_B)$ varies 
smoothly, $\hat r_{\text{IS}}$ will
have a small variance. With our choice of 
partitioning in~(\ref{eqn:PartG}) and~(\ref{eqn:PartL}),
we expect to observe a small variance 
if the Ising model is in a strong (negative) external magnetic 
field, see Appendix~I. 

The choice of partitioning on the dual graph is 
arbitrary, as long as $\tilde \X_B$ can be computed as linear combinations
of $\tilde \X_A$. Our choice of partitioning is suitable for models in a strong 
external magnetic field. 
Depending on the values of the model parameters 
and their spatial distributions,
different choices of partitioning will yield schemes with different dynamics. 

If the model is not in a very strong external field, 
we can consider
applying annealed importance sampling~\cite{NealIS:2001}, \cite{LoMo:IT2013}; see Appendix~II. 
For models in a weak external field, the efficiency of the importance sampling algorithm 
on the dual factor graph should be 
compared to the efficiency of Monte Carlo methods applied directly to the original factor 
graph, as in Figs.~\ref{fig:2DGrid} and~\ref{fig:2DGridOrig}.


We can design a uniform sampling scheme
by drawing each $x_{A, k}^{(\ell)}$ uniformly and independently 
from $\calX$, and by applying 
\begin{equation} 
\label{eqn:EstU}
\hat r_{\text{Unif}} = 
\frac{|\calX|^{|\B|}}{L} \sum_{\ell = 1}^L \Gamma(\tilde \x_A^{(\ell)})\Lambda(\tilde \x_B^{(\ell)})
\end{equation}

It is easy to verify that, $\E[\, \hat r_{\text{Unif}}\,] = Z_\text{d}$.
\vspace{0.67mm}

It must be emphasized that, the cost of generating $\tilde \x_{A}^{(\ell)}$ using the importance sampling 
scheme (i.e., drawing an independent sample according to~(\ref{eqn:AuxDist})) is virtually the same as the cost of
generating a sample with uniform sampling (i.e., drawing a sample uniformly and independently in the state space).

The efficiency of the uniform sampling 
algorithm and the importance sampling scheme will be close if the model is 
at very low temperature, i.e., $J_k$ is very large. However,
for a wider range of model parameters, importance 
sampling outperforms uniform sampling, as will be illustrated in our numerical 
experiments in Section~\ref{sec:Num}; see Appendix~I.
Applying uniform sampling and Gibbs sampling in the dual domain
to 2D Ising models in the absence of an external field are discussed in~\cite{MoLo:ISIT2013}.


\section{The $q$-State Potts Model in an External Magnetic Field}
\label{sec:Potts}

In a 2D $q$-state Potts model, $\calX = \{0, 1, \ldots, q-1\}$,
where $q$ is an integer greater than
or equal to 2. The energy 
of a configuration $\x$ is given by the Hamiltonian
\begin{equation}
\label{eqn:HamiltonianP}
\mathcal{H}_{\text{Potts}}(\x) \eqdef -\!\!\!\sum_{\text{$(k,\ell)\in \B$}}\!\!\!J_{k, \ell}
[x_k = x_{\ell}]
- \sum_{m = 1}^N H_m [x_m = 0] 
\end{equation}

Here, $J_{k, \ell}$ controls the strength of
the interaction between adjacent variables $(x_k, x_{\ell})$ and
$H_m$ corresponds to the presence of an external 
magnetic field\footnote{Our definition of the Hamiltonian~(\ref{eqn:HamiltonianP}) is 
based on the assumption that 
$H_m$ applies 
only if $x_m = 0$. The Hamiltonian of the Potts model can be defined
in other ways, e.g., the external field can apply when
$x_m = 1$ or when $x_m$ is in more than one sate.}.
For $q = 2$, the Potts model is equivalent to the Ising 
model.

Similar to the 2D Ising model in Section~\ref{sec:Ising}, for each adjacent pair $(x_k, x_\ell)$, we let
\begin{equation}
\label{eqn:PottsA}
\kappa_{k, \ell}(x_k, x_{\ell}) 
= e^{J_{k, \ell}[x_k = x_{\ell}]}
\end{equation}
and for each $x_m$
\begin{equation}
\label{eqn:PottsH}
\tau_{m}(x_m) = e^{H_m [x_m = 0]}
\end{equation}


The Forney factor graph of a 2D $q$-state Potts model is 
similar to the factor graph in~\Fig{fig:2DGrid},
where
the unlabeled normal-size 
boxes represent factors as in~(\ref{eqn:PottsA}), and the
small boxes represent factors as in~(\ref{eqn:PottsH}).

In the dual Forney graph, the XOR factors are 
as in~(\ref{eqn:XOR}), where $\oplus$ denotes addition in GF($q$). 
The 1D DFT of factors as in~(\ref{eqn:PottsH}) are 
\begin{equation} 
\label{eqn:PottsKernelDual2}
\lambda_{m}(\tilde x_m) = \left\{ \begin{array}{lr}
     e^{H_m} + q -1, & \text{if $\tilde x_m = 0$} \\
     e^{H_m} - 1, & \text{otherwise,} 
\end{array} \right.
\end{equation}
and the 2D DFT of~(\ref{eqn:PottsA}) will have the following form

\begin{equation} 
\label{eqn:PottsKernelDual}
\gamma_{k}(\tilde x_k, \tilde x_{\ell}) = \left\{ \begin{array}{ll}
      q(e^{J_{k, {\ell}}} + q-1), & \text{if $\tilde x_k = \tilde x_{\ell}= 0$} \\
      q(e^{J_{k, {\ell}}} -1), & \text{if $\tilde x_k \oplus \tilde x_{\ell}= 0$} \\
      0,                    & \text{otherwise,}
  \end{array} \right.
\end{equation}
where $\oplus$ denotes addition in GF($q$).

The corresponding dual Forney factor graph is 
shown in~\Fig{fig:2DGridD}, where the small 
boxes represent factors as in~(\ref{eqn:PottsKernelDual2}) and 
the unlabeled normal-size boxes 
represent factors as in~(\ref{eqn:PottsKernelDual}).

By adding extra XOR factors on each bond,
we can obtain the modified dual Forney factor graph of the $q$-state
Potts model 
with factors attached to each equality constraint as

\begin{equation} 
\label{eqn:PottsKernelDualS}
\gamma_{k}(\tilde x_k) = \left\{ \begin{array}{ll}
      q(e^{J_k} + q-1), & \text{if $\tilde x_k = 0$} \\
      q(e^{J_k} -1), & \text{otherwise.}
  \end{array} \right.
\end{equation}

\begin{figure}
\setlength{\unitlength}{0.91mm}
\centering
\begin{picture}(77,72.5)(0,0)
\small
\put(0.5,60.5){\framebox(3,3){$+$}}
 \put(3.5,60.5){\line(4,-3){4}}
 \put(7.5,54.5){\framebox(3,3){}}
\put(3.5,62){\line(1,0){8.5}}        
\put(12,60){\framebox(4,4){$=$}}
\put(16,62){\line(1,0){3}}
\put(22,62){\line(1,0){2.5}}
\put(19,60.5){\framebox(3,3){$+$}}
\put(24.5,60.5){\framebox(3,3){$+$}}
 \put(27.5,60.5){\line(4,-3){4}}
 \put(31.5,54.5){\framebox(3,3){}}
\put(27.5,62){\line(1,0){8.5}}       
\put(36,60){\framebox(4,4){$=$}}
\put(40,62){\line(1,0){3}}
\put(46,62){\line(1,0){2.5}}
\put(43,60.5){\framebox(3,3){$+$}}
\put(48.5,60.5){\framebox(3,3){$+$}}
 \put(51.5,60.5){\line(4,-3){4}}
 \put(55.5,54.5){\framebox(3,3){}}
\put(51.5,62){\line(1,0){8.5}}       
\put(60,60){\framebox(4,4){$=$}}
\put(64,62){\line(1,0){3}}
\put(70,62){\line(1,0){2.5}}
\put(67,60.5){\framebox(3,3){$+$}}
\put(72.5,60.5){\framebox(3,3){$+$}}
 \put(75.5,60.5){\line(4,-3){4}}
 \put(79.5,54.5){\framebox(3,3){}}
\put(2,54){\line(0,1){6.5}}
\put(0,50){\framebox(4,4){$=$}}
\put(0.5,45){\framebox(3,3){$+$}}
\put(2,50){\line(0,-1){2}}
\put(2,45){\line(0,-1){1.5}}
\put(26,54){\line(0,1){6.5}}
\put(24,50){\framebox(4,4){$=$}}
\put(24.5,45){\framebox(3,3){$+$}}
\put(26,50){\line(0,-1){2}}
\put(26,45){\line(0,-1){1.5}}
\put(50,54){\line(0,1){6.5}}
\put(48,50){\framebox(4,4){$=$}}
\put(48.5,45){\framebox(3,3){$+$}}
\put(50,50){\line(0,-1){2}}
\put(50,45){\line(0,-1){1.5}}
\put(74,54){\line(0,1){6.5}}
\put(72,50){\framebox(4,4){$=$}}
\put(72.5,45){\framebox(3,3){$+$}}
\put(74,50){\line(0,-1){2}}
\put(74,45){\line(0,-1){1.5}}
\put(0.5,40.5){\framebox(3,3){$+$}}
 \put(3.5,40.5){\line(4,-3){4}}
 \put(7.5,34.5){\framebox(3,3){}}
\put(3.5,42){\line(1,0){8.5}}
\put(12,40){\framebox(4,4){$=$}}
\put(16,42){\line(1,0){3}}
\put(22,42){\line(1,0){2.5}}
\put(19,40.5){\framebox(3,3){$+$}}
\put(24.5,40.5){\framebox(3,3){$+$}}
 \put(27.5,40.5){\line(4,-3){4}}
 \put(31.5,34.5){\framebox(3,3){}}
\put(27.5,42){\line(1,0){8.5}}
\put(36,40){\framebox(4,4){$=$}}
\put(40,42){\line(1,0){3}}
\put(46,42){\line(1,0){2.5}}
\put(43,40.5){\framebox(3,3){$+$}}
\put(48.5,40.5){\framebox(3,3){$+$}}
 \put(51.5,40.5){\line(4,-3){4}}
 \put(55.5,34.5){\framebox(3,3){}}
\put(51.5,42){\line(1,0){8.5}}
\put(60,40){\framebox(4,4){$=$}}
\put(64,42){\line(1,0){3}}
\put(70,42){\line(1,0){2.5}}
\put(67,40.5){\framebox(3,3){$+$}}
\put(72.5,40.5){\framebox(3,3){$+$}}
 \put(75.5,40.5){\line(4,-3){4}}
 \put(79.5,34.5){\framebox(3,3){}}
\put(2,34){\line(0,1){6.5}}
\put(0,30){\framebox(4,4){$=$}}
\put(0.5,25){\framebox(3,3){$+$}}
\put(2,30){\line(0,-1){2}}
\put(2,25){\line(0,-1){1.5}}
\put(26,34){\line(0,1){6.5}}
\put(24,30){\framebox(4,4){$=$}}
\put(24.5,25){\framebox(3,3){$+$}}
\put(26,30){\line(0,-1){2}}
\put(26,25){\line(0,-1){1.5}}
\put(50,34){\line(0,1){6.5}}
\put(48,30){\framebox(4,4){$=$}}
\put(48.5,25){\framebox(3,3){$+$}}
\put(50,30){\line(0,-1){2}}
\put(50,25){\line(0,-1){1.5}}
\put(74,34){\line(0,1){6.5}}
\put(72,30){\framebox(4,4){$=$}}
\put(72.5,25){\framebox(3,3){$+$}}
\put(74,30){\line(0,-1){2}}
\put(74,25){\line(0,-1){1.5}}
\put(0.5,20.5){\framebox(3,3){$+$}}
 \put(3.5,20.5){\line(4,-3){4}}
 \put(7.5,14.5){\framebox(3,3){}}
\put(3.5,22){\line(1,0){8.5}}
\put(12,20){\framebox(4,4){$=$}}
\put(16,22){\line(1,0){3}}
\put(22,22){\line(1,0){2.5}}
\put(19,20.5){\framebox(3,3){$+$}}
\put(24.5,20.5){\framebox(3,3){$+$}}
 \put(27.5,20.5){\line(4,-3){4}}
 \put(31.5,14.5){\framebox(3,3){}}
\put(27.5,22){\line(1,0){8.5}}
\put(36,20){\framebox(4,4){$=$}}
\put(40,22){\line(1,0){3}}
\put(46,22){\line(1,0){2.5}}
\put(43,20.5){\framebox(3,3){$+$}}
\put(48.5,20.5){\framebox(3,3){$+$}}
 \put(51.5,20.5){\line(4,-3){4}}
 \put(55.5,14.5){\framebox(3,3){}}
\put(51.5,22){\line(1,0){8.5}}
\put(60,20){\framebox(4,4){$=$}}
\put(64,22){\line(1,0){3}}
\put(70,22){\line(1,0){2.5}}
\put(67,20.5){\framebox(3,3){$+$}}
\put(72.5,20.5){\framebox(3,3){$+$}}
 \put(75.5,20.5){\line(4,-3){4}}
 \put(79.5,14.5){\framebox(3,3){}}
\put(2,14){\line(0,1){6.5}}
\put(0,10){\framebox(4,4){$=$}}
\put(0.5,5){\framebox(3,3){$+$}}
\put(2,10){\line(0,-1){2}}
\put(2,5){\line(0,-1){1.5}}
\put(26,14){\line(0,1){6.5}}
\put(24,10){\framebox(4,4){$=$}}
\put(24.5,5){\framebox(3,3){$+$}}
\put(26,10){\line(0,-1){2}}
\put(26,5){\line(0,-1){1.5}}
\put(50,14){\line(0,1){6.5}}
\put(48,10){\framebox(4,4){$=$}}
\put(48.5,5){\framebox(3,3){$+$}}
\put(50,10){\line(0,-1){2}}
\put(50,5){\line(0,-1){1.5}}
\put(74,14){\line(0,1){6.5}}
\put(72,10){\framebox(4,4){$=$}}
\put(72.5,5){\framebox(3,3){$+$}}
\put(74,10){\line(0,-1){2}}
\put(74,5){\line(0,-1){1.5}}
\put(0.5,0.5){\framebox(3,3){$+$}}
 \put(3.5,0.5){\line(4,-3){4}}
 \put(7.5,-5.5){\framebox(3,3){}}
\put(3.5,2){\line(1,0){8.5}}
\put(12,0){\framebox(4,4){$=$}}
\put(16,2){\line(1,0){3}}
\put(22,2){\line(1,0){2.5}}
\put(19,0.5){\framebox(3,3){$+$}}
\put(24.5,0.5){\framebox(3,3){$+$}}
 \put(27.5,0.5){\line(4,-3){4}}
\put(31.5,-5.5){\framebox(3,3){}}
\put(27.5,2){\line(1,0){8.5}}
\put(36,0){\framebox(4,4){$=$}}
\put(40,2){\line(1,0){3}}
\put(46,2){\line(1,0){2.5}}
\put(43,0.5){\framebox(3,3){$+$}}
\put(48.5,0.5){\framebox(3,3){$+$}}
 \put(51.5,0.5){\line(4,-3){4}}
 \put(55.5,-5.5){\framebox(3,3){}}
\put(51.5,2){\line(1,0){8.5}}
\put(60,0){\framebox(4,4){$=$}}
\put(64,2){\line(1,0){3}}
\put(70,2){\line(1,0){2.5}}
\put(67,0.5){\framebox(3,3){$+$}}
\put(72.5,0.5){\framebox(3,3){$+$}}
 \put(75.5,0.5){\line(4,-3){4}}
 \put(79.5,-5.5){\framebox(3,3){}}
 
\put(8,63){\pos{bc}{$\tilde X_1$}}
\put(32,63){\pos{bc}{$\tilde X_2$}}

\put(14,64){\line(0,1){2}}
\put(38,64){\line(0,1){2}}
\put(62,64){\line(0,1){2}}
\put(12,66){\framebox(4,4){$$}}
\put(36,66){\framebox(4,4){$$}}
\put(60,66){\framebox(4,4){$$}}
\put(14,44){\line(0,1){2}}
\put(38,44){\line(0,1){2}}
\put(62,44){\line(0,1){2}}
\put(12,46){\framebox(4,4){$$}}
\put(36,46){\framebox(4,4){$$}}
\put(60,46){\framebox(4,4){$$}}
\put(14,24){\line(0,1){2}}
\put(38,24){\line(0,1){2}}
\put(62,24){\line(0,1){2}}
\put(12,26){\framebox(4,4){$$}}
\put(36,26){\framebox(4,4){$$}}
\put(60,26){\framebox(4,4){$$}}
\put(14,4){\line(0,1){2}}
\put(38,4){\line(0,1){2}}
\put(62,4){\line(0,1){2}}
\put(12,6){\framebox(4,4){$$}}
\put(36,6){\framebox(4,4){$$}}
\put(60,6){\framebox(4,4){$$}}
\put(0,52){\line(-1,0){2}}
\put(24,52){\line(-1,0){2}}
\put(48,52){\line(-1,0){2}}
\put(72,52){\line(-1,0){2}}
\put(-6,50){\framebox(4,4){$$}}
\put(18,50){\framebox(4,4){$$}}
\put(42,50){\framebox(4,4){$$}}
\put(66,50){\framebox(4,4){$$}}
\put(0,32){\line(-1,0){2}}
\put(24,32){\line(-1,0){2}}
\put(48,32){\line(-1,0){2}}
\put(72,32){\line(-1,0){2}}
\put(-6,30){\framebox(4,4){$$}}
\put(18,30){\framebox(4,4){$$}}
\put(42,30){\framebox(4,4){$$}}
\put(66,30){\framebox(4,4){$$}}
\put(0,12){\line(-1,0){2}}
\put(24,12){\line(-1,0){2}}
\put(48,12){\line(-1,0){2}}
\put(72,12){\line(-1,0){2}}
\put(-6,10){\framebox(4,4){$$}}
\put(18,10){\framebox(4,4){$$}}
\put(42,10){\framebox(4,4){$$}}
\put(66,10){\framebox(4,4){$$}}
\end{picture}
\vspace{3ex}
\caption{\label{fig:2DGridPotts}%
The modified dual Forney factor graph of the 2D Potts model in 
an external field. The small 
boxes represent factors as in~(\ref{eqn:PottsKernelDual2}),
the unlabeled normal-size boxes 
represent factors as in~(\ref{eqn:PottsKernelDualS}), 
and boxes containing 
$+$ symbols represent XOR factors as in~(\ref{eqn:XOR}), where $\oplus$ denotes 
addition in GF($q$).
}
\end{figure}
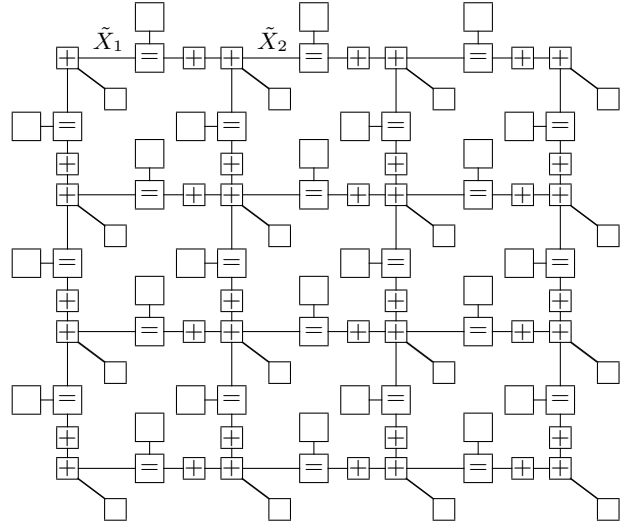

Here, $J_k$ is 
the coupling parameter associated with each bond.
\Fig{fig:2DGridPotts} shows the corresponding modified dual Forney factor 
graph with factors 
as in~(\ref{eqn:PottsKernelDual2}) 
and~(\ref{eqn:PottsKernelDualS})\footnote{These extra XOR 
factors (sign inverters in~\cite{Forney:01, Forney:11}) are not required in~\Fig{fig:2DGridDM}, as 
variables are binary, and $\oplus$ denotes 
addition in GF(2).}.

We consider ferromagnetic Potts models
in a positive external field, 
characterized by $J_k > 0$ and $H_m > 0$, respectively. Therefore, all the factors in
(\ref{eqn:PottsKernelDual2}) and~(\ref{eqn:PottsKernelDualS}) will
be positive. 

The importance sampling scheme  
can be generalized to the $q$-state Potts model with little effort.
We are not going to repeat the complete scheme here. 
We only point out that following the set-up of Section~\ref{sec:IS}, we have
\begin{equation}
Z_q = q^{2|\B|}\text{exp}\Big(\sum_{k \in \B} J_k\Big)
\end{equation}

To draw $\tilde \x_A^{(\ell)}$ according to $q(\tilde \x_A)$, 
we apply 
\vspace{0.5ex}

\begin{itemize}
\itemsep1.7pt
\item[] \emph{draw} $u_1^{(\ell)}, u_2^{(\ell)}, \ldots, u_{|\B|}^{(\ell)}\overset{\text{i.i.d.}}{\sim} \, \mathcal{U}[0,1]$
\item[] {\bf for} $k = 1$ {\bf to} $|\B|$
\item [] \hspace{4.5mm} {\bf if} $u_k^{(\ell)} < \dfrac{1+(q-1)e^{-J_k}}{q}$
\item [] \hspace{10mm}  $\tilde x_{A,k}^{(\ell)} = 0$
\item [] \hspace{4.5mm} {\bf else}
\item [] \hspace{10mm}  \emph{draw} $\tilde x_{A,k}^{(\ell)}$ \emph{randomly from} $\{1,2,\ldots,q-1\}$
\item [] \hspace{4.5mm} {\bf end if}
\item[] {\bf end for}
\end{itemize}
\vspace{0.5ex}


The estimator in~(\ref{eqn:EstR}) is expected to 
have a small variance if the Potts model is in a strong (positive) external
magnetic field, see~(\ref{eqn:PottsKernelDual2}).

If the Potts model is at very low temperature, one might instead use the following 
algorithm based on uniform sampling

\vspace{0.5ex}

\begin{itemize}
\itemsep1.8pt
\item[] \emph{draw} $u_1^{(\ell)}, u_2^{(\ell)}, \ldots, u_{|\B|}^{(\ell)}\overset{\text{i.i.d.}}{\sim} \, \mathcal{U}[0,1]$
\item[] {\bf for} $k = 1$ {\bf to} $|\B|$
\item [] \hspace{4.5mm} $\tilde x_{A, k}^{(\ell)} = \lfloor\, q\cdot u_k^{(\ell)} \rfloor$
\item[] {\bf end for}
\end{itemize}
\vspace{0.5ex}

%
%
Here, $\lfloor \cdot \rfloor$ denotes the floor function~\cite{Knuth:92}.

The importance sampling and uniform sampling algorithms become
equivalent when $J_k \to \infty$ for $k \in \B$.

For known analytical results regarding the Potts 
model see~\cite[Chapter 12]{Baxter07} and~\cite{GJ:2012}.

\begin{figure}[t!]
\centering
\includegraphics[width=1.0\linewidth]{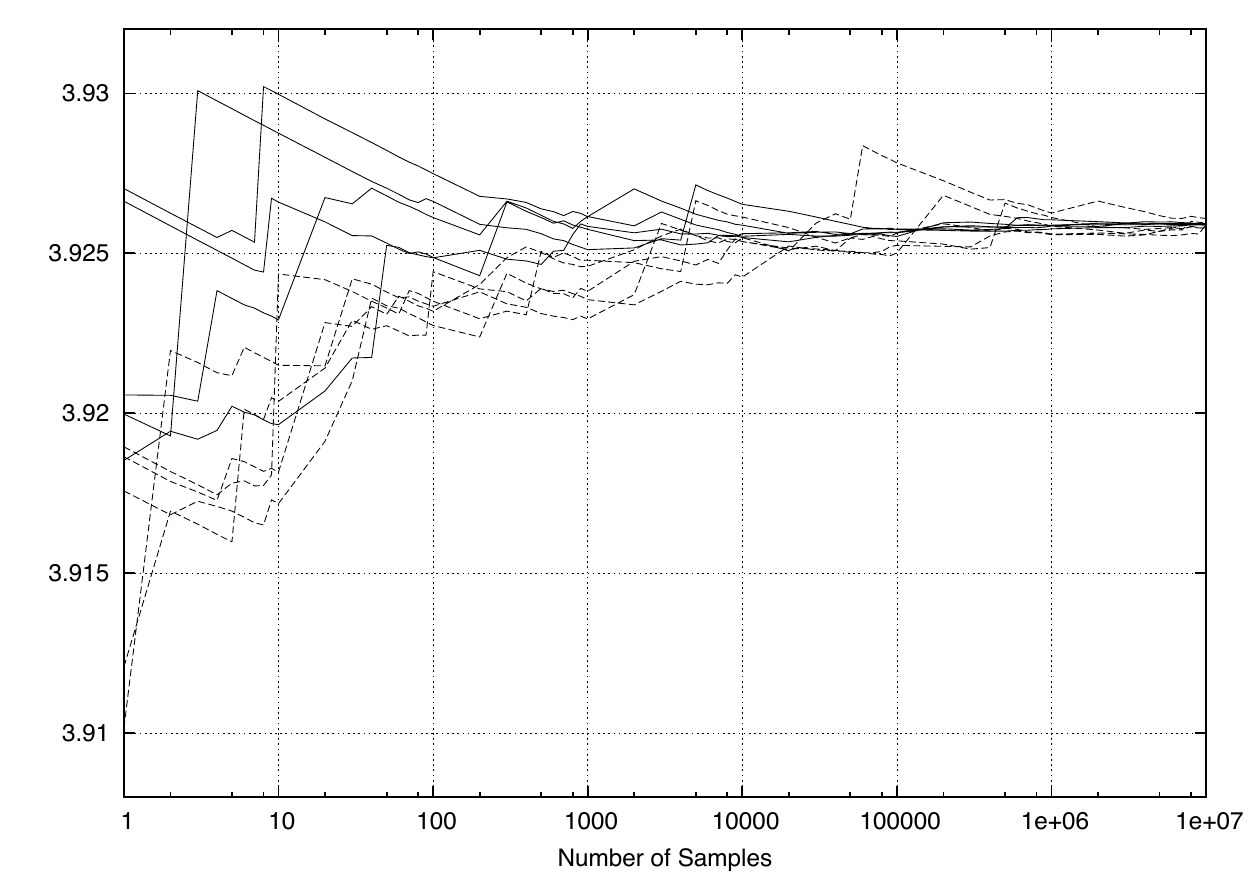}
\caption{\label{fig:FerIsing1}%
Estimated free energy per site vs.\ the number of samples
for a $30\times 30$ Ising model, with
$J\sim\calU[1.3, 1.5]$ and $H\sim\calU[-1.25, -1.0]$ (strong field). 
The plot shows five different sample paths obtained from importance sampling (solid lines) and five different sample paths obtained from uniform sampling (dashed lines)
in the dual factor graph.}
\vspace{3.0ex}
\includegraphics[width=1.0\linewidth]{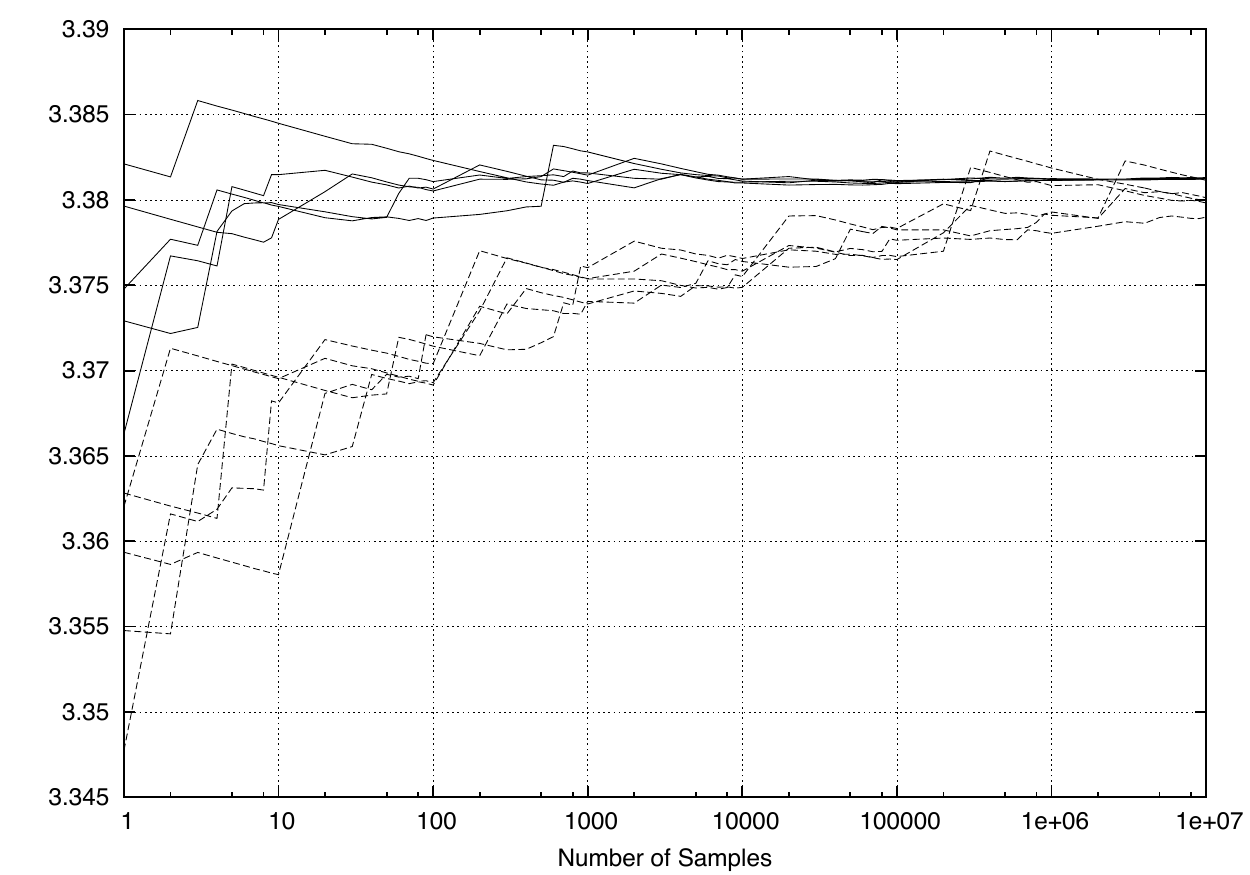}
\caption{\label{fig:FerIsing2}%
Everything as in Fig.~\ref{fig:FerIsing1}, but with $J\sim\calU[0.75, 1.5]$.}
\end{figure}

\section{Numerical Experiments}
\label{sec:Num}
%

We apply the importance sampling and the uniform sampling
schemes of Section~\ref{sec:IS}
to estimate the free energy~(\ref{eqn:FreeEnergy}) per 
site, i.e., $\frac{1}{N}\ln Z$, of 
the 2D and 3D ferromagnetic Ising model and the 2D ferromagnetic Potts 
model.

\begin{figure}[t!]
\includegraphics[width=1.0\linewidth]{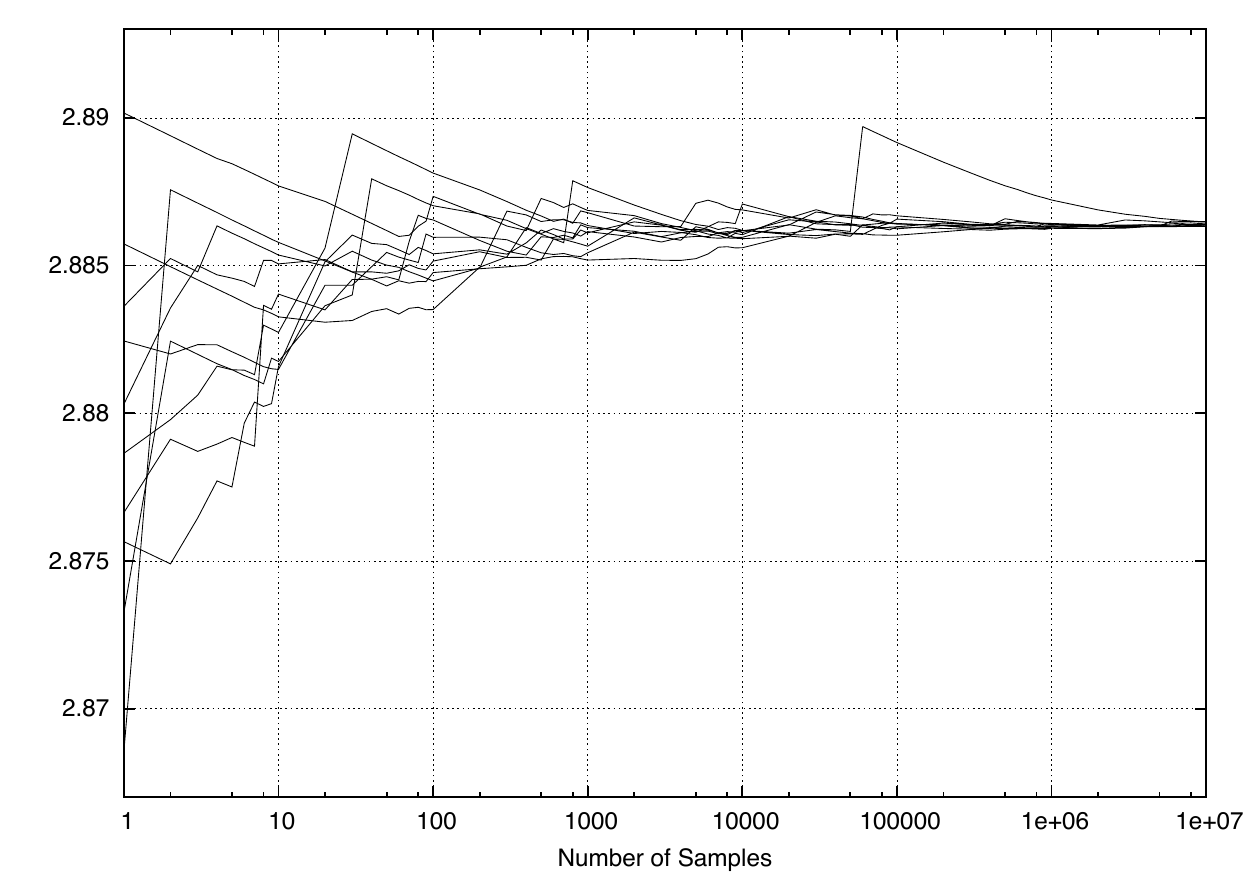}
\caption{\label{fig:FerIsing3}%
Estimated free energy per site vs.\ the number of samples
for a $30\times 30$ Ising model, with $J\sim\calU[0.25, 1.5]$ and $H\sim\calU[-1.25, -1.0]$. 
The plot shows ten different sample paths obtained from importance sampling
in the dual factor graph.}
\vspace{3.0ex}
\includegraphics[width=1.0\linewidth]{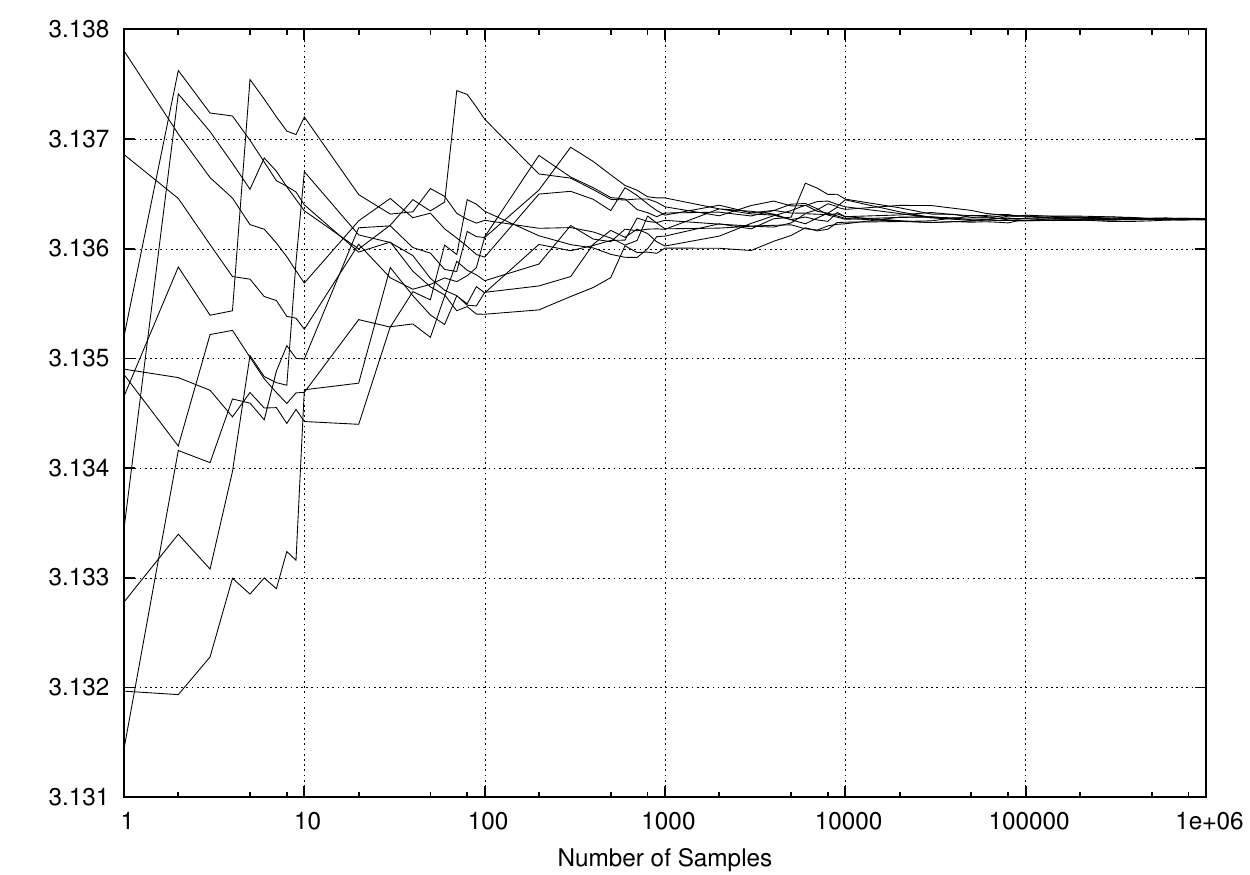}
\caption{\label{fig:FerIsing44}%
Everything as in Fig.~\ref{fig:FerIsing3}, but with $H \sim\calU[-1.5, -1.25]$.}
\end{figure}





In Section~\ref{sec:NumIsing}, we consider 2D 
ferromagnetic Ising models in an external field 
with spatially varying model parameters.
We recall from Section~\ref{sec:Ising} that the value of $Z$ is invariant under the
change of sign of the external field;
we set $H_m <  0$ to make all the factors 
as in~(\ref{eqn:IsingKernelDual2}) positive.
In Section~\ref{sec:NumIsing3D}, we consider 3D Ising models defined
on a cubic lattice. 2D 
ferromagnetic Potts models with spatially varying couplings
in a positive external magnetic field, $H_m >  0$, are considered
in Section~\ref{sec:NumPotts}.

All simulation results show $\frac{1}{N}\ln Z$ vs.\ the number
of samples for ``one instance" of the models with periodic
boundary conditions, where to create periodic boundary conditions
we need to add extra edges (with appropriate factors) 
to connect the 
sites on opposite sides of the boundary. In this case $|\B| = 2N$. 
For different realizations 
of the Ising model, 
we will report the histogram of the estimated free energy per site.

\subsection{2D Ising model} 
\label{sec:NumIsing}

We consider 2D Ising models of 
size $N = 30\times 30$
in all the experiments. 

In our first two experiments we
set $H_{m} \overset{\text{i.i.d.}}{\sim} \calU[-1.25, -1.0]$.
The coupling parameters 
are set to $J_{k} \overset{\text{i.i.d.}}{\sim} \calU[1.3, 1.5]$ 
in the first experiment and to $J_{k} \overset{\text{i.i.d.}}{\sim} \calU[0.75, 1.5]$ 
in the second experiment. 
Simulation results for one instance of the model obtained from importance 
sampling (solid lines) and uniform sampling (dashed lines) in the dual factor 
graph are shown in Figs.~\ref{fig:FerIsing1} and \ref{fig:FerIsing2}.
The estimated free energy per site is about $3.926$ and $3.381$, respectively.

For very large coupling parameters (corresponding to models at very low 
temperature), convergence of uniform sampling is comparable to the 
convergence of the importance
sampling algorithm, see~\Fig{fig:FerIsing1}. 
However, in Fig~\ref{fig:FerIsing2} we observe that uniform sampling has issues 
with slow convergence 
for a wider range of coupling parameters,
while the proposed importance sampling scheme performs
well in all the ranges.

\begin{figure}[t!]
\includegraphics[width=1.01\linewidth]{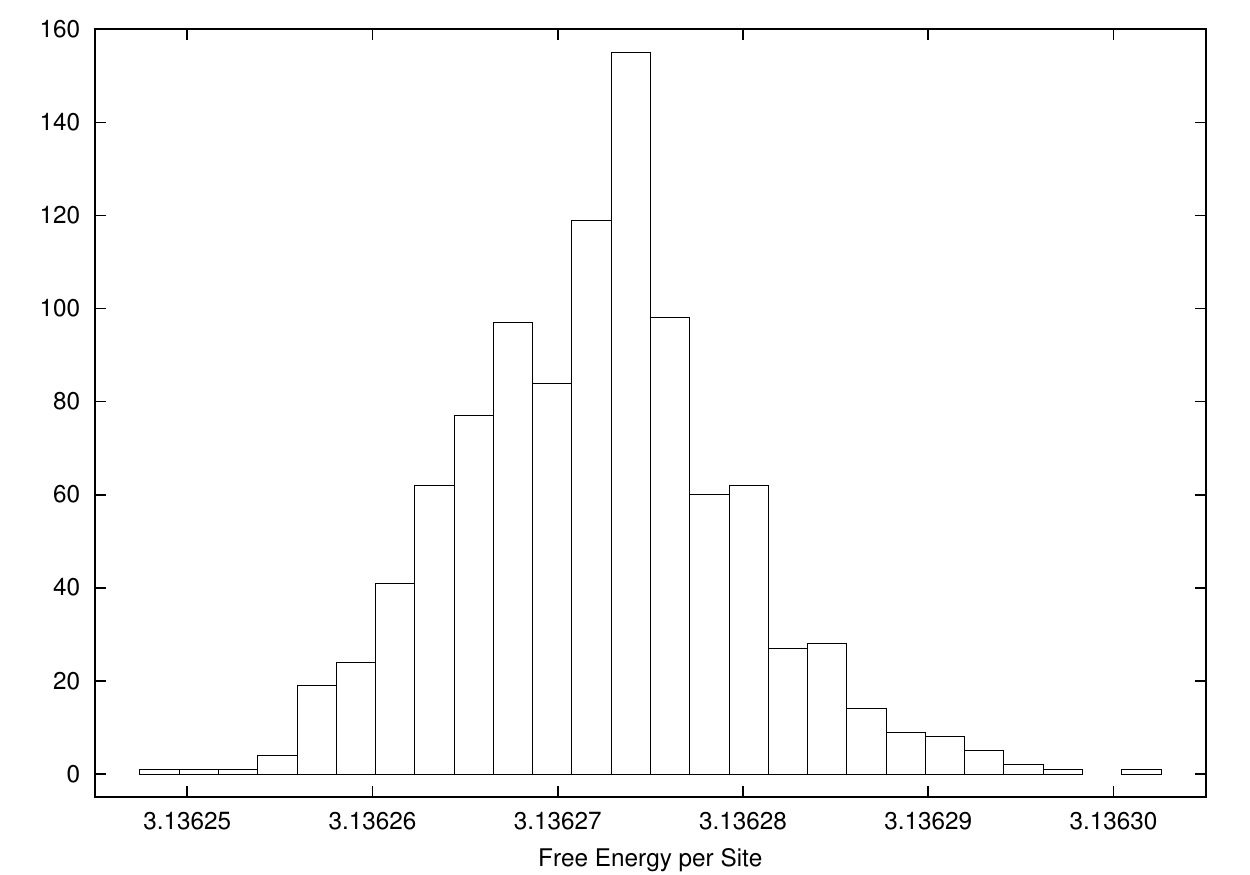}
\caption{\label{fig:Hist2D}%
Histogram for $1000$ realizations
of the estimated free energy per site for a $30\times 30$ Ising model, 
with $J\sim\calU[0.25, 1.5]$ and $H\sim\calU[-1.5, -1.25]$.}
\end{figure}

In our second two experiments we
set $J_{k} \overset{\text{i.i.d.}}{\sim} \calU[0.25, 1.5]$.
In the third experiment, we set $H_{m} \overset{\text{i.i.d.}}{\sim} \calU[-1.25, -1.0]$.
Fig.~\ref{fig:FerIsing3} shows simulation 
results for one instance of the model obtained from importance sampling, where the estimated free energy per site 
is about $2.886$. We set $H_{m} \overset{\text{i.i.d.}}{\sim} \calU[-1.5, -1.25]$ in the last
experiment. For one instance of the model, the estimated $\frac{1}{N}\ln Z$ from~Fig.~\ref{fig:FerIsing44} is 
about $3.1362$. We observe that convergence of the importance sampling
algorithm improves as $|H|$ becomes larger; see Appendix~I.

Fig.~\ref{fig:Hist2D} shows the histogram of the estimated free energy per site for 
$1000$ realizations of a $30\times 30$ Ising model, 
with $J\overset{\text{i.i.d.}}{\sim} \calU[0.25, 1.5]$ and $H \overset{\text{i.i.d.}}{\sim} \calU[-1.5, -1.25]$.
Using the {\bf R} package, the fitted normal distribution to Fig.~\ref{fig:Hist2D} has mean
equal to $3.136272$ and the standard deviation equal to $7.401838\times 10^{-6}$.

\subsection{3D Ising model} 
\label{sec:NumIsing3D}

\begin{figure}[t!]
\includegraphics[width=1.01\linewidth]{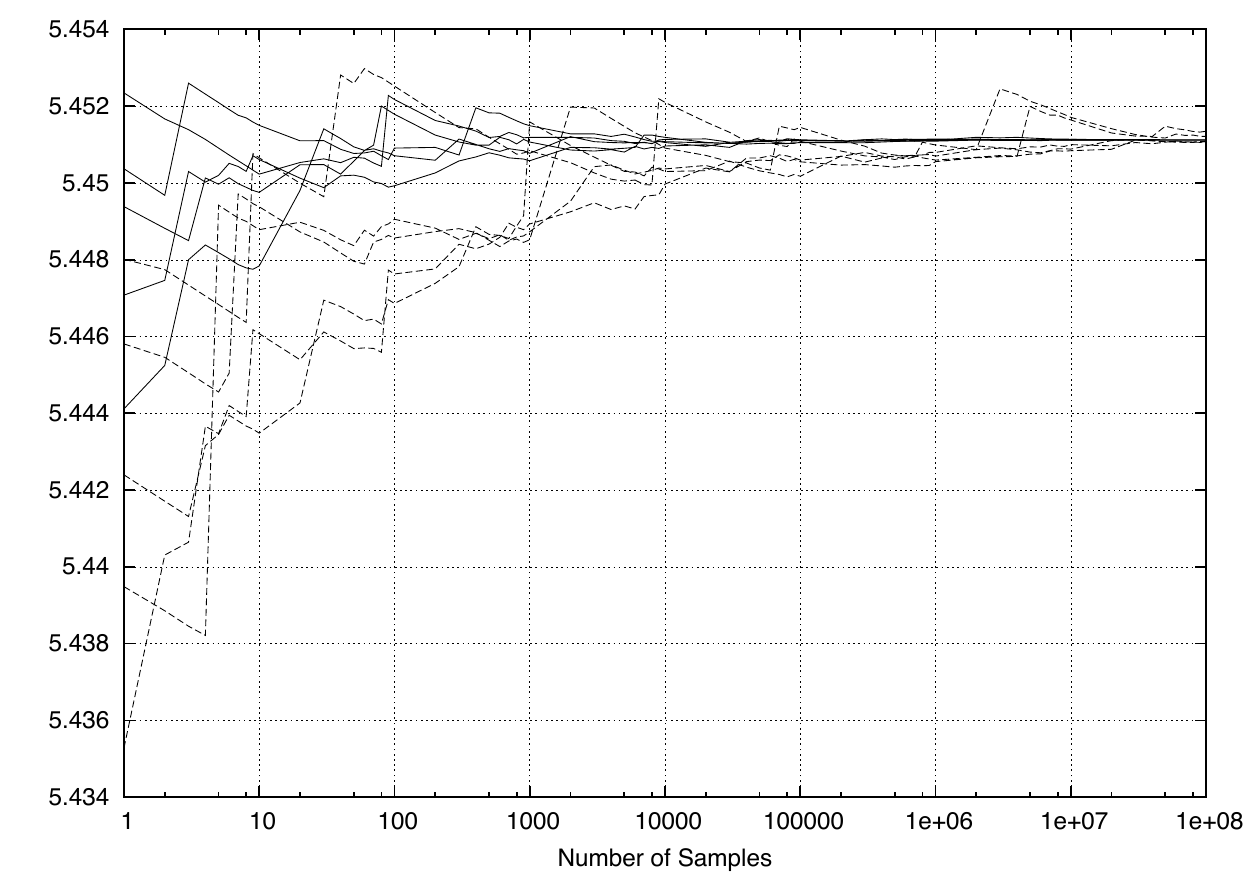}
\caption{\label{fig:FerIsing3D}%
Estimated free energy per site vs.\ the number of samples
for a $10\times 10\times 10$ ferromagnetic Ising model in an external field with periodic
boundary conditions, with
$J\sim\calU[1.0, 2.0]$ and $H = -1.5$. 
The plot shows five different sample paths obtained from importance 
sampling (solid lines) and five different sample paths obtained from uniform 
sampling (dashed lines)
on the dual factor graph.}
\end{figure}

The method can be applied to ferromagnetic 3D Ising models in an
external field.
In a model of size $N = 10\times 10\times 10$,
we set $J_{k,\ell} \overset{\text{i.i.d.}}{\sim} \calU[1.0, 2.0]$ 
and $H = -1.5$. For one instance of the Ising model, simulation results 
for one instance of the model obtained from importance 
sampling (solid lines) and uniform sampling (dashed lines) on the dual 
factor graph are shown in \Fig{fig:FerIsing3D}, where the estimated 
free energy per site, i.e., $\frac{1}{N}\ln Z$, is about $5.451$.

\begin{figure}[t!]
\includegraphics[width=1.01\linewidth]{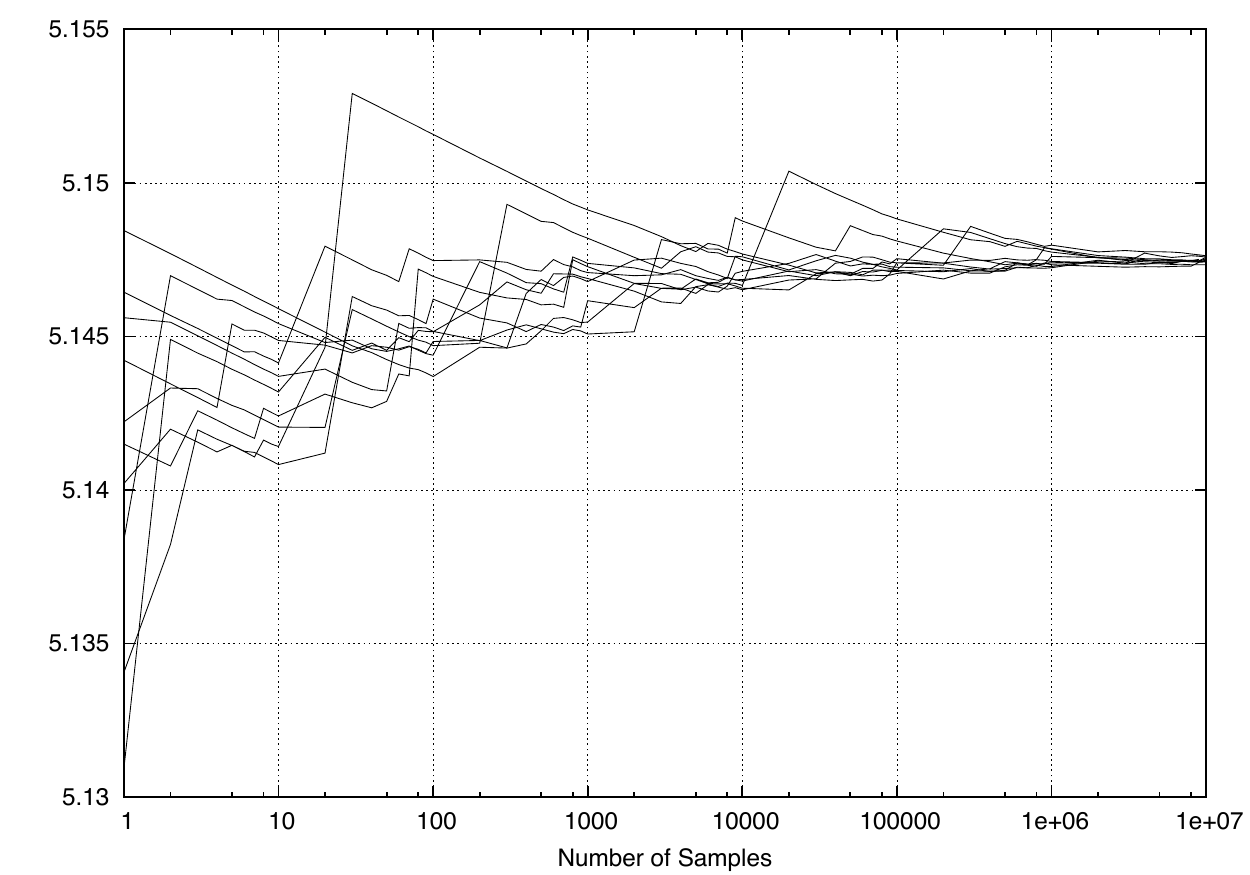}
\caption{\label{fig:FerPotts}%
Estimated free energy per site vs.\ the number of samples
for a $30\times 30$ ferromagnetic 3-state Potts model in an external field with periodic
boundary conditions, with
$J\sim\calU[0.25, 2.5]$ and $H\sim\calU[2.25, 2.5]$ (strong field). 
The plot shows ten different sample paths obtained from importance sampling
on the dual factor graph.
}
\end{figure}

\subsection{2D Potts model} 
\label{sec:NumPotts}

We consider a 2D $3$-state 
Potts model of size $N = 30\times 30$ in an external field, with 
$J_{k,\ell} \overset{\text{i.i.d.}}{\sim} \calU[0.25, 2.5]$ 
and $H_{m} \overset{\text{i.i.d.}}{\sim} \calU[2.25, 2.5]$. For one instance of the model, 
\Fig{fig:FerPotts} shows simulation results 
obtained from importance sampling on 
the dual factor for one instance of the model. The estimated 
free energy per site is about $5.147$.

\section{Conclusion}

An importance sampling scheme on the dual Forney factor graph was proposed %
to estimate the 
partition function of 2D and 3D
ferromagnetic Ising and 2D 
ferromagnetic $q$-state Potts models, when the 
models are in 
the presence of
an external magnetic field. 
We described a method to partition the variables on the dual graph and
introduced an auxiliary importance sampling distribution accordingly. 
The
method can efficiently compute an estimate of the partition function under a wide range 
of model parameters, in particular (with our choice of partitioning), when 
the models are in a strong external field.
Depending on the values of the model parameters 
and their spatial distributions,
different choices of partitioning yield schemes with different 
convergence properties.

\section*{Appendix I\\}

For simplicity, we assume that the coupling parameter and the external field are constant, 
denoted by $J$ and $H$, respectively.
It is numerically advantageous to replace each factor as in~(10)
in the dual factor graph by
\begin{equation} 
\label{eqn:IsingKerneltanh1}
\lambda(\tilde x_m) = (\tanh |H|)^{\tilde x_m} 
\end{equation}
and each factor as in~(12) by 
\begin{equation} 
\label{eqn:IsingKerneltanh2}
\gamma(\tilde x_k) = (\tanh J)^{\tilde x_k}
\end{equation}

The required scale factor $S$ to recover $Z_\text{d}$ can be easily computed by multiplying all 
the local scale factors as
\begin{equation} 
\label{eqn:DDRatio}
S = (4\cosh J)^{|\B|}(2\cosh H)^N
\end{equation}

Note that, $\lim_{t \to \infty} \tanh t = 1$, therefore
in a strong external magnetic field (i.e., large $|H|$) and at low temperature 
(i.e., large $J$), $\tanh |H|$
and $\tanh J$ both tend to constant, which gives reasons 
for the fast convergence of uniform sampling in this case. 
In our importance sampling scheme, independent samples 
are drawn according to $q(\tilde \x_A)$ in~(\ref{eqn:AuxDist}), thus the only 
requirement to achieve fast convergence is
having a strong
external field. 

Indeed, in the dual domain, convergence 
of the importance sampling algorithm improves as $|H|$ becomes larger, and
convergence of uniform sampling improves as $J$ and $|H|$ both become larger, which is in sharp 
contrast with Monte Carlo methods in the original domain.

We analyze the variance of the importance sampling
algorithm for estimating the partition function of ``finite-size" 2D 
models more rigorously. Let $p_\text{d}(\tilde \x_A)$ denote the global probability mass function
in the dual Forney factor graph. Notice that $p_\text{d}(\cdot)$ and $q(\cdot)$ in~(\ref{eqn:AuxDist}) are both 
defined in the same 
configuration space $\calX^{|\B|}$.

We can therefore write $p_\text{d}(\cdot)$ as a function of $\tilde \x_A$, as
\begin{IEEEeqnarray}{r,C,l}
\label{eqn:GlobDist2}
p_{\text{d}}(\tilde \x_A) & = & \frac{\Gamma(\tilde \x_A)\Lambda(\tilde \x_B)}{Z_\text{d}},  \qquad\, \forall\, \tilde \x_A \in \calX^{|\B|} \\
                                    & = &  \frac{Z_q}{Z_\text{d}}q(\tilde \x_A)\Lambda(\tilde \x_B),   \qquad\, \forall\, \tilde \x_A \in \calX^{|\B|}
\end{IEEEeqnarray}
where $\tilde \x_B$ is a linear combination of $\tilde \x_A$. 
%
%


The variance of $\hat Z_{\text{IS}}$ in~(\ref{eqn:EstR}) can be computed as 
\begin{IEEEeqnarray}{r,C,l}
\V [\, \hat Z_{\text{IS}} \,] &=&\E\big[\hat Z_{\text{IS}}^2\big] - 
\big(\!\E\big[ \hat Z_{\text{IS}}\big]\big)^2 \\
&=& \frac{1}{L} \big(Z_q^2\cdot\E_q[\,\Lambda^2(\tilde \X_B)\,] - Z_{\text{d}}^2 \big) \label{eqn:VARu}
\end{IEEEeqnarray}

To remind ourselves that $\hat Z_{\text{IS}}$ is a function of $L$, we 
write it as $\hat Z_{\text{IS}}(L)$.
We then rewrite~(\ref{eqn:VARu}) as
\begin{IEEEeqnarray}{r,C,l}
    \frac{L}{Z_{\text{d}}^2} \V [\, \hat Z_{\text{IS}}(L) \,]  
    & = &  \Big(\frac{Z_q}{Z_{\text{d}}}\Big)^2\E_q[\, \Lambda^2(\tilde \X_B)\,] - 1 \\
    & = & \sum_{\tilde \x_A} \frac{p^2_{\text{d}}(\tilde \x_A)}{q(\tilde \x_A)} - 1 \\
    & = & \chi^2\big(p_{\text{d}}(\tilde \x_A), q(\tilde \x_A)\big) \label{eqn:VarBound}
\end{IEEEeqnarray}
where $\chi^2(\cdot, \cdot)$ denotes the chi-squared divergence, which is always non-negative, 
with equality to zero if and only if its two arguments are equal~\cite[Chapter 4]{CS:04}.

%
%

In the limit $|H| \to \infty$, we have (see also~(\ref{eqn:IsingKerneltanh1}))
\begin{IEEEeqnarray}{r,C,l}
\lim_{|H| \to \infty}  p_{\text{d}}(\tilde \x_A) & = & q(\tilde \x_A),  \qquad\, \forall\, \tilde \x_A \in \calX^{|\B|}
\end{IEEEeqnarray}

Hence
\begin{IEEEeqnarray}{r,C,l}
\lim_{|H| \to \infty}  \chi^2\big(p_{\text{d}}(\tilde \x_A), q(\tilde \x_A)\big) & = & 0 \label{eqn:ChiLim}
\end{IEEEeqnarray}

We conclude that $Z_{\text{d}}$ can be estimated efficiently via the proposed
importance sampling estimator when the model is in a strong external field.

\section*{Appendix II\\ Annealed Importance Sampling in the Dual Forney Factor Graph}

We briefly explain how to employ annealed importance sampling in the dual factor 
graph to estimate the partition function of the 2D Ising model, when the model is not in a very strong 
external field.

For simplicity, we assume that the coupling parameter and the external field are both constant. 
The partition function is thus denoted by $Z_{\text{d}}(J, |H|)$. 
We express $Z_{\text{d}}(J, |H|)$ using a sequence of intermediate partition functions 
by varying $|H|$ in $V$ levels as
\begin{equation}
\label{eqn:AIS}
Z_{\text{d}}(J, |H|) = Z_{\text{d}}(J, |H|^{\alpha_V})\prod_{v = 0}^{V-1} \frac{Z_{\text{d}}(J, |H|^{\alpha_v})}{Z_{\text{d}}(J, |H|^{\alpha_{v+1}})}
\end{equation}

Here, unlike typical annealing strategies in the original domain, 
$(\alpha_0, \alpha_1, \ldots, \alpha_V)$ is an 
increasing 
sequence with $1 = \alpha_0 < \alpha_1 < \cdots < \alpha_V$.

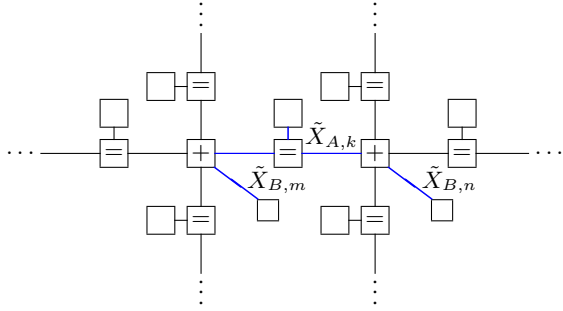
\begin{figure}[t]
\centering
\setlength{\unitlength}{0.89mm}
\begin{picture}(76, 37)(0, 0)
\small
\put(-1,11.9){\pos{bc}{$\ldots$}}
\put(2,12){\line(1,0){9}}
\put(11,10){\framebox(4,4){$=$}}
\put(15,12){\line(1,0){9}}
\put(13,14){\line(0,1){2}}
\put(11,16){\framebox(4,4){$$}}
\put(37,10){\framebox(4,4){$=$}}
\put(37,16){\framebox(4,4){$$}}

\put(37.4,6){\pos{bc}{$\tilde X_{B, m}$}}
\put(63.1,6){\pos{bc}{$\tilde X_{B, n}$}}
\put(54,12){\line(1,0){9}}
\put(67,12){\line(1,0){8}}
\put(78,11.9){\pos{bc}{$\ldots$}}
\put(24,10){\framebox(4,4){$+$}}
 \put(34.5,2){\framebox(3,3){}}
\put(45.5,12.5){\pos{bc}{$\tilde X_{A,k}$}}
\put(50,10){\framebox(4,4){$+$}}
 \put(60.5,2){\framebox(3,3){}}
\put(63,10){\framebox(4,4){$=$}}
\put(24,2){\line(-1,0){2}}
\put(50,2){\line(-1,0){2}}
\put(24,22){\line(-1,0){2}}
\put(50,22){\line(-1,0){2}}
\put(65,14){\line(0,1){2}}
\put(24,20){\framebox(4,4){$=$}}
\put(50,20){\framebox(4,4){$=$}}
\put(24,0){\framebox(4,4){$=$}}
\put(50,0){\framebox(4,4){$=$}}
\put(18,0){\framebox(4,4){$$}}
\put(44,0){\framebox(4,4){$$}}
\put(63,16){\framebox(4,4){$$}}
\put(18,20){\framebox(4,4){$$}}
\put(44,20){\framebox(4,4){$$}}

\put(26,31){\pos{bc}{$\vdots$}}
\put(26,-10.5){\pos{bc}{$\vdots$}}
\put(26,24){\line(0,1){6}}
\put(26,20){\line(0,-1){6}}
\put(52,24){\line(0,1){6}}
\put(52,20){\line(0,-1){6}}
\put(52,31){\pos{bc}{$\vdots$}}
\put(52,-10.5){\pos{bc}{$\vdots$}}
\put(26,4){\line(0,1){6}}
\put(26,0){\line(0,-1){6}}
\put(52,4){\line(0,1){6}}
\put(52,0){\line(0,-1){6}}
\color{blue}
\put(28,10){\line(4,-3){6.5}}
 \put(54,10){\line(4,-3){6.5}}
\put(28,12){\line(1,0){9}}
\put(41,12){\line(1,0){9}}
\put(39,14){\line(0,1){2}}
\put(28,10){\line(4,-3){6.5}}
 \put(54,10){\line(4,-3){6.5}}
\put(28,12){\line(1,0){9}}
\put(41,12){\line(1,0){9}}
\put(39,14){\line(0,1){2}}
\end{picture}
\vspace{6.0ex}
\caption{\label{fig:2DGridGibbs}%
In Gibbs sampling in the dual Forney factor graph of a 2D Ising model,
changing $\tilde x_{A,k}$ involves changing $\tilde x_{B, m}$ and $\tilde x_{B, n}$ at 
the same time.
}
\end{figure}

If $\alpha_V$ is large enough, $Z_{\text{d}}(J, |H|^{\alpha_V})$ can be estimated efficiently via the proposed 
importance 
sampling scheme. As for the intermediate steps, a sampling 
technique that leaves the target distribution invariant (e.g., Metropolis algorithms or Gibbs sampling), is 
required at each level. These intermediate target probability distributions correspond to
the intermediate partition functions.

Here, we explain how Gibbs 
sampling can be applied to draw samples in the dual Forney graph of the 2D Ising 
model.
Gibbs sampling is performed on $\tilde \X_A$ (i.e., the variables on the bonds).
Each iteration $\ell$, consists of visiting the bonds sequentially, from $k = 1$ to $|\B|$, in the 
dual Forney factor graph, and updating $\tilde x^{(\ell)}_{A,k}$ using Gibbs sampling.

%
%

In updating $\tilde x^{(\ell)}_{A,k}$, the value of $\tilde \x_{A}\backslash \tilde x_{A,k}$ 
is irrelevant. Therefore, changing the value of $\tilde x_{A,k}$ will only involve a simultaneous change in the 
values of $\tilde x_{B, m}$ and 
$\tilde x_{B, n}$. These variables (edges) are marked blue in~\Fig{fig:2DGridGibbs}.
At the end of the $\ell$-th iteration (i.e., when $k = |\B|$), the Gibbs sampling algorithm will generate 
$\tilde \x_{A}^{(\ell)}$ and $\tilde \x_{B}^{(\ell)}$. 
The algorithm
is iterated for a predetermined number of times at each level; the last generated sample is 
usually used as the initial state of the Gibbs sampler at the 
next level.

The number of levels $V$ should be sufficiently large to
ensure that intermediate target distributions are close enough 
and estimating $Z_{\text{d}}(J, |H|^{\alpha_V})$ 
is feasible; see~\cite{NealIS:2001}.


\section*{Acknowledgements}

 
The author would like to thank Hans-Andrea Loeliger, Pascal Vontobel,  and Justin Dauwels 
for their 
comments that greatly improved the presentation of this paper. 
%


\newcommand{\IT}{IEEE Trans.\ Information Theory}
\newcommand{\CASI}{IEEE Trans.\ Circuits \& Systems~I}
\newcommand{\COM}{IEEE Trans.\ Comm.}
\newcommand{\COMLet}{IEEE Commun.\ Lett.}
\newcommand{\COMMag}{IEEE Communications Mag.}
\newcommand{\ETT}{Europ.\ Trans.\ Telecomm.}
\newcommand{\SPMag}{IEEE Signal Proc.\ Mag.}
\newcommand{\ProcIEEE}{Proceedings of the IEEE}

\end{document}